\documentclass[longauth]{aa}  
\usepackage[varg]{txfonts}

\usepackage{graphicx}	
\usepackage{amsmath}	
\usepackage{amssymb}	
\usepackage{hyperref}   
\usepackage{upgreek}    
\usepackage{soul}       

\usepackage[dvipsnames]{xcolor}
\setstcolor{red}
\usepackage{ulem}
\usepackage[many]{tcolorbox}

\usepackage{natbib}
\bibpunct{(}{)}{;}{a}{}{,} 
\usepackage{balance}

\usepackage[para,online,flushleft]{threeparttable}
\usepackage{array,multirow,booktabs}
\usepackage{bigstrut}

\usepackage{braket}

\usepackage{nicefrac}

\newtcolorbox{cross}{blank,breakable,parbox=false,
  overlay={\draw[red,line width=1pt] (interior.south west)--(interior.north east);
    \draw[red,line width=1pt] (interior.north west)--(interior.south east);}}

\newcommand*{\earth}{{\oplus}}

\makeatletter
\renewcommand*\aa@pageof{, page \thepage{} of \pageref*{LastPage}}
\makeatother

\begin{document}

   \title{A warm super-Neptune around the G-dwarf star TOI-1710 \\ revealed with \textit{TESS}, SOPHIE and HARPS-N}

   \author{P.-C. König\inst{1,2}
      \and M.~Damasso\inst{3}
      \and G.~Hébrard\inst{2,4}
      \and L.~Naponiello\inst{5,6}
      \and P.~Cortés-Zuleta\inst{7}
      \and K.~Biazzo\inst{8}%
      \and N.~C.~Santos\inst{9,10}
      \and A.~S.~Bonomo\inst{3}
      \and A.~Lecavelier~des~Étangs\inst{2}
      \and L.~Zeng\inst{11,12}
      \and S.~Hoyer\inst{7}
      \and A.~Sozzetti\inst{3}
      \and L.~Affer\inst{13}
      \and J.~M.~Almenara\inst{14}
      \and S.~Benatti\inst{13}
      \and A.~Bieryla\inst{11}
      \and I.~Boisse\inst{7}
      \and X.~Bonfils\inst{14}
      \and W.~Boschin\inst{15,16,17}
      \and A.~Carmona\inst{14}
      \and R.~Claudi\inst{18}
      \and K.~A.~Collins\inst{11}
      \and S.~Dalal\inst{2}
      \and M.~Deleuil\inst{7}
      \and X.~Delfosse\inst{14}
      \and O.~D.~S.~Demangeon\inst{9,10}
      \and S.~Desidera\inst{18}
      \and R.~F.~Díaz\inst{19}
      \and T.~Forveille\inst{14}
      \and N.~Heidari\inst{20,21,7}
      \and G. A. J.~Hussain\inst{1,22}
      \and J.~Jenkins\inst{23}
      \and F.~Kiefer\inst{2,24}
      \and G.~Lacedelli\inst{25,18}
      \and D.~W.~Latham\inst{11}
      \and L.~Malavolta\inst{18,25}
      \and L.~Mancini\inst{5,3,26}
      \and E.~Martioli\inst{2,27}
      \and G.~Micela\inst{13}
      \and P.~A.~Miles-Páez\inst{1}
      \and C.~Moutou\inst{28}
      \and D.~Nardiello\inst{7,18}
      \and V.~Nascimbeni\inst{18}
      \and M.~Pinamonti\inst{3}
      \and G.~Piotto\inst{18,25}
      \and G.~Ricker\inst{29}
      \and R.~P.~Schwarz\inst{30}
      \and S.~Seager\inst{31,29,32}
      \and R.~G.~Stognone\inst{33}
      \and P.~A.~Str{\o}m\inst{2,34,35}
      \and R.~Vanderspek\inst{29}
      \and J.~Winn\inst{36}
      \and J.~Wittrock\inst{37}
      }

   \institute{European Southern Observatory, Karl-Schwarzschild-Straße 2, D-85748 Garching, Germany\\
              \email{\href{mailto:pkonig@eso.org}{pkonig@eso.org}}
         \and Institut d'astrophysique de Paris, CNRS, UMR 7095 \& Sorbonne Université, UPMC Paris 6, 98bis Bd Arago, 75014 Paris, France
         \and INAF -- Osservatorio Astrofisico di Torino, via Osservatorio 20, I-10025 Pino Torinese, Italy
         \and Observatoire de Haute-Provence, CNRS, Universit\'{e} d’Aix-Marseille, 04870 Saint-Michel-l'Observatoire, France
         \and Department of Physics, University of Rome Tor Vergata, Via della Ricerca Scientifica 1, I-00133 Rome, Italy
         \and Department of Physics and Astronomy, University of Florence, Via Sansone 1, I-50019, Sesto Fiorentino (FI), Italy
         \and Aix-Marseille Université, CNRS, CNES, LAM, Marseille, France
         \and INAF -- Osservatorio Astronomico di Roma, via Frascati 33, I-00040, Monte Porzio Catone (RM), Italy
         \and Instituto de Astrofísica e Ciências do Espaço, Universidade do Porto, CAUP, Rua das Estrelas, 4150-762 Porto, Portugal
         \and Departamento de Física e Astronomia, Faculdade de Ciências, Universidade do Porto, Rua do Campo Alegre, 4169-007 Porto, Portugal
         \and Harvard-Smithsonian Center for Astrophysics, 60 Garden Street, Cambridge, MA 02138, USA
         \and Department of Earth and Planetary Sciences, Harvard University, 20 Oxford Street, Cambridge, MA 02138, USA
         \and INAF -- Osservatorio Astronomico di Palermo, Piazza del Parlamento, 1, I-90134, Palermo, Italy
         \and Université Grenoble Alpes, CNRS, IPAG, 38000 Grenoble, France
         \and Fundación Galileo Galilei -- INAF, Rambla José Ana Fernandez Pérez 7, 38712 Bre{\~n}a Baja, TF, Spain
         \and Instituto de Astrofísica de Canarias, C/Vía Láctea s/n, 38205 La Laguna, TF, Spain
         \and Departamento de Astrofísica, Universidad de La Laguna, 38206 La Laguna, TF, Spain
         \and INAF -- Osservatorio Astronomico di Padova, Vicolo dell’Osser-vatorio 5, I-35122, Padova, Italy
         \and International Center for Advanced Studies and ICIFI (CONICET), ECyT-UNSAM, Campus Miguelete, 25 de Mayo y Francia, (1650) Buenos Aires, Argentina
         \and Department of Physics, Shahid Beheshti University, Tehran, Iran
         \and Laboratoire J.-L. Lagrange, OCA, Université de Nice-Sophia Antipolis, CNRS, Campus Valrose, 06108 Nice Cedex 2, France
         \and European Space Agency (ESA), European Space Research and Technology Centre (ESTEC), Keplerlaan 1, 2201 AZ Noordwijk, The Netherlands
         \and NASA Ames Research Center, Moffett Field, CA, 94035, USA
         \and LESIA, Observatoire de Paris, Université PSL, CNRS, Sorbonne Université, Université de Paris, Meudon, France
         \and Dip. di Fisica e Astronomia Galileo Galilei, Università di Padova, Vicolo dell’Osservatorio 2, I-35122 Padova, Italy
         \and Max Planck Institute for Astronomy Königstuhl 17, 69117, Heidelberg, Germany
         \and Laboratório Nacional de Astrofísica, Rua Estados Unidos 154, 37504-364, Itajubá-MG, Brazil
         \and IRAP, Université de Toulouse, CNRS, UPS, CNES, F-31400 Toulouse, France
         \and Department of Physics and Kavli Institute for Astrophysics and Space Research, MIT, Cambridge, MA 02139, USA
         \and Patashnick Voorheesville Observatory, Voorheesville, NY 12186, USA
         \and Department of Earth, Atmospheric, and Planetary Sciences, MIT, Cambridge, MA 02139, USA
         \and Department of Aeronautics and Astronautics, MIT, Cambridge, MA 02139, USA
         \and Dipartimento di Fisica, Universit\`a degli Studi di Torino, via Pietro Giuria 1, I-10125, Torino, Italy
         \and Department of Physics, University of Warwick, Coventry CV4 7AL, UK
         \and Centre for Exoplanets and Habitability, University of Warwick, Gibbet Hill Road, Coventry CV4 7AL, UK
         \and Department of Astrophysical Sciences, Peyton Hall, 4 Ivy Lane, Princeton, NJ 08544, USA
         \and George Mason University, 4400 University Drive, Fairfax, VA, 22030 USA
         }

   \date{Received 24 December 2021 / Accepted 14 April 2022}

 
  \abstract{We report the discovery and characterization of the transiting extrasolar planet TOI-1710\:b. It was first identified as a promising candidate by the \textit{Transiting Exoplanet Survey Satellite} (\textit{TESS}). Its planetary nature was then established with SOPHIE and HARPS-N spectroscopic observations via the radial-velocity method. The stellar parameters for the host star are derived from the spectra and a joint Markov chain Monte-Carlo (MCMC) adjustment of the spectral energy distribution and evolutionary tracks of TOI-1710. A joint MCMC analysis of the \textit{TESS} light curve and the radial-velocity evolution allows us to determine the planetary system properties. From our analysis, TOI-1710\:b is found to be a massive warm super-Neptune ($M_{\rm p}=28.3\:\pm\:4.7\:{\rm M}_{\rm \earth}$ and $R_{\rm p}=5.34\:\pm\:0.11\:{\rm R}_{\rm \earth}$) orbiting a G5V dwarf star ($T_{\rm eff}=5665\pm~55\mathrm{K}$) on a nearly circular 24.3-day orbit ($e=0.16\:\pm\:0.08$). The orbital period of this planet is close to the estimated rotation period of its host star $P_{\rm rot}=22.5\pm2.0~\mathrm{days}$ and it has a low Keplerian semi-amplitude $K=6.4\pm1.0~\mathrm{m\:s^{-1}}$; we thus performed additional analyses to show the robustness of the retrieved planetary parameters. With a low bulk density of $1.03\pm0.23~\mathrm{g\:cm^{-3}}$ and orbiting a bright host star ($J=8.3$, $V=9.6$), TOI-1710\:b is one of the best targets in this mass-radius range (near the Neptunian desert) for atmospheric characterization via transmission spectroscopy, a key measurement in constraining planet formation and evolutionary models of sub-Jovian planets.}
   
            
   \maketitle
%

\section{Introduction}
\label{sec:introduction}

Extrasolar planets that transit their host stars are particularly interesting to study. 
Observing these objects in both spectroscopy and photometry allows us to determine their orbital and physical properties, including mass and radius. 
According to the \texttt{exoplanet.eu} archive \citep{2011A&A...532A..79S}, about 1000 exoplanets have such a double identification, the majority of which were first identified from photometric surveys and then characterized with radial-velocity (RV) follow-up. 
Spectroscopic observations of planetary candidates revealed by photometry are used to establish or reject their planetary nature, namely by measuring the mass of the transiting bodies using the RV method \citep[e.g.,][]{2007ARA&A..45..397U,2011arXiv1109.2497M,2014A&A...572A..93H,2021MNRAS.500.5088C}. 
They are also used to confirm the orbital period and measure other
orbital parameters of the companion, such as the eccentricity, and to characterize its host star. 
Additionally, upon further exploration, long-term RV follow-up can also reveal other potential planets (including non-transiting planets) within the same system \citep[e.g.,][]{2009A&A...506..303Q}.

Studying the population of planets and their fundamental properties in a statistical manner is crucial to better understanding the dominating processes of planet formation and evolution. Following on from the Convection, Rotation, and planetary Transits (CoRoT) space telescope, the first mission capable of detecting small transiting exoplanets \citep{2009A&A...506..411A}, the Kepler mission is the most successful survey of transiting exoplanets to date \citep{2010Sci...327..977B,2014PASP..126..398H}, and has significantly impacted the field with the detection of more than 2300
exoplanets \citep[see NASA Exoplanet Archive,][]{2013PASP..125..989A}.
Nevertheless, due to the faintness of the stars targeted by Kepler, only a small fraction are suitable for RV follow-up. 

The space observatory \textit{Transiting Exoplanet Survey Satellite} (\textit{TESS}) was designed to survey 85\% of the sky, observing successively multiple sectors of the sky, specifically searching for transiting exoplanets around bright nearby stars \citep{2015JATIS...1a4003R}. 
During the first three years of its mission, in which \textit{TESS} searched for planets alternating between the southern and northern ecliptic hemispheres, nearly 5000 candidates were detected, of which 170 have been characterized as exoplanets. It has now started Years~4 and 5 of observations.

One of the key features in the exoplanet population is known as the hot-Neptune desert, which corresponds to a lack of Neptune-mass planets close to their host stars \citep{2011ApJ...727L..44S} likely due to photoevaporation and/or tidal disruption phenomena \citep{2007A&A...461.1185L,2016A&A...589A..75M}. 
Detecting and characterizing new planets within this range has become crucial to understanding this desert; \textit{TESS} has been able to find some of the rare planets populating it, for example the hot Neptune LTT\:9779\:b \citep{2020NatAs...4.1148J}. 
The designs of the two ground-based spectrographs SOPHIE \citep{2006tafp.conf..319B,2008SPIE.7014E..0JP,2011SPIE.8151E..15P} and HARPS-N \citep{2012SPIE.8446E..1VC, 2014SPIE.9147E..8CC} are appropriate to study a range of planets of this size and mass. For instance, the warm Neptunes Gl\:378b \citep{2019A&A...625A..18H} and HD\:164595b \citep{2015A&A...581A..38C} were discovered with SOPHIE and the two sub-Saturns K2-79b and K2-222b \citep{2022AJ....163...41N} and the pair of hot Neptunes TOI-942\:b and TOI-942\:c \citep{2021A&A...645A..71C} with HARPS-N. 
For this study we used the two instruments to discover the warm super-Neptune hosted by the \textit{TESS} Object of Interest (TOI) 1710, exploring the upper edge of the desert.

TOI-1710 is a nearby ($d=81\:\mathrm{pc}$) G5V dwarf star (see Table~\ref{tab:star} for a full summary of the stellar properties). 
TOI-1710\:b was first identified as an exoplanetary candidate in the \textit{TESS} Data Validation Report\footnote{Mikulski Archive for Space Telescopes, \textit{TESS} Science Processing Operations Center Pipeline, NASA Ames Research Center, \texttt{\href{https://mast.stsci.edu/portal/Mashup/Clients/Mast/Portal.html}{https://mast.stsci.edu}}} of 2 August 2020 by photometric observations from \textit{TESS} in Sectors 19, 20, and 26, revealing three distinct transits.
A fourth transit was then observed one year later in Sector~40, confirming the expected period.
The detection and characterization are also based on RV follow-up observations with SOPHIE and HARPS-N asserting the planetary origin of the transits detected in the \textit{TESS} light curve (LC) and precisely determining the properties of the planetary system.

TOI-1710\:b is particularly favorable for atmospheric characterization.
Beyond planet detection and the progress in planetary demographics, remarkable advances have been made through spectroscopic observations toward understanding and characterizing the atmospheres of these alien worlds. 
Detailed characterization of a given planet's atmosphere provides the essential insight for exploring the interplay between its initial composition \citep[e.g.,][]{2011Natur.469...64M,2012A&A...547A.111M}, chemistry \citep[e.g.,][]{2010ApJ...709.1396F,2015ApJ...804...61B}, dynamics and circulation \citep[e.g.,][]{2009ApJ...699..564S}, winds \citep[e.g.,][]{2020A&A...633A..86S}, disequilibrium processes \citep[e.g.,][]{2013ApJ...763...25M,2013ApJ...779....3L}, and atmospheric escape \citep[e.g.,][]{2004A&A...418L...1L,2016A&A...591A.121B,2018Sci...362.1384A,2018Sci...362.1388N}, shedding light on the underlying physical processes in these distant worlds. 
In recent years, high-resolution transmission spectroscopy with ground-based observatories has revealed numerous atomic \citep[e.g.,][]{2008A&A...487..357S,2020ApJ...888L..13T,2021ApJ...919L..15D} and molecular \citep[e.g.,][]{2010Natur.465.1049S} species in exoplanet atmospheres. 
Comprehensive overviews of exoplanet atmosphere observations and models describe the highlights in the field \citep[][]{2010ARA&A..48..631S,2014Natur.513..345B,2015PASP..127..941C,2016SSRv..205..285M,2018arXiv180408149F}. \\

The paper is structured as follows. In Sect.~\ref{sec:observations} we give an overview of the data acquisition, the reduction methods, and the main features of the collected data. We then present in Sect.~\ref{sec:star_prop} the analysis of the host star's fundamental properties and give in Sect.~\ref{sec:transit_keplerian_characterization} the detail of the modeling of the LC and the RV time series along with our main results. Finally, in Sect.~\ref{sec:discussion_conclusions} we discuss our method and the system, and present our conclusions.


\section{Observations and data reduction}
\label{sec:observations}


\subsection{Photometric identification with \textsc{TESS}}
\label{sec:photometric_identification_TESS}

\begin{figure*}
\centering
\includegraphics[trim=0.24cm 0.cm 0.0cm 0.cm, clip=true, scale=0.57]{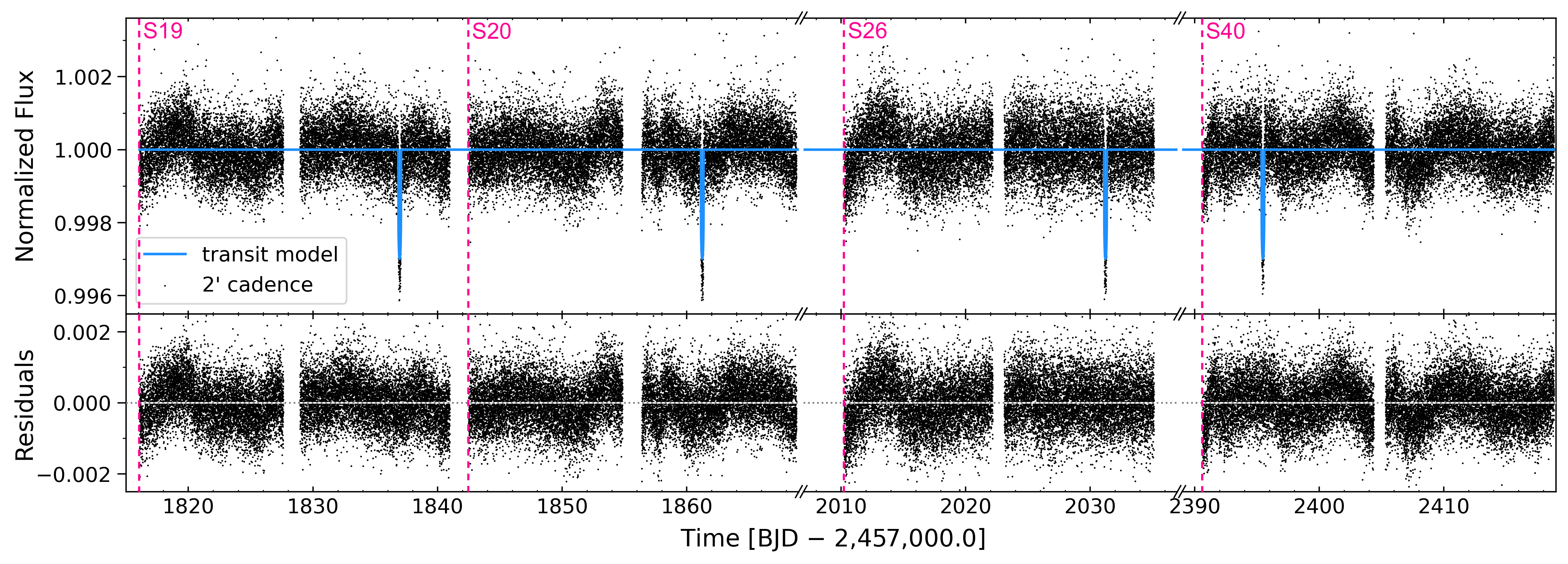}
\caption{LC of TOI-1710 collected by \textit{TESS} with a 2 min cadence in Sectors~19, 20, 26, and 40 (upper panel). The blue line shows the transit profile of the best-fit model presented in this study (Sect.~\ref{sec:transit_keplerian_characterization}) and is used to compute the residuals in the bottom panel. The four \textit{TESS} sectors are delimited by vertical fuchsia dashed lines.}\label{fig:LC}
\end{figure*}

\textit{TESS} observes the sky in sectors measuring $24^\circ \times 96^\circ$ with a focal ratio of $f/1.4$ and a broadband filter range from 600 to 1000~nm. 
The two-minute cadence data were reduced with the Science Processing Operations Center (SPOC) pipeline \citep{2016SPIE.9913E..3EJ} adapted from the pipeline for the Kepler mission at the NASA Ames Research Center in order to produce calibrated pixels and LCs and to search for signatures of transiting planets. In this work we use the systematic error-corrected Presearch Data Conditioned Simple Aperture Photometry (PDC-SAP) LC \citep{2012PASP..124..985S,2014PASP..126..100S,2012PASP..124.1000S} provided by the \textit{TESS} pipeline and retrieved through the Python package \texttt{lightkurve} \citep{2018ascl.soft12013L}. We removed any low-quality data point (flagged by the pipeline) as well as any others at more than five times the standard deviation of the data set; this sigma-clipping was applied for less than 0.1\% of the data. 

The \textit{TESS} photometric measurements of TOI-1710 were acquired between 28 November 2019 and 23 July 2021.
The final LC consists of 70\:875 two-minute exposures, with a median uncertainty of $604\:\mathrm{ppm}$, spanning 603 days with two main gaps, between Sectors 21--25 and between Sectors 27--39 (i.e., between 20 January 2020 and 9 June 2020, and between 6 July 2020 and 24June 2021, respectively. The normalized LC is shown in Fig.~\ref{fig:LC}~(top); images with the \textit{TESS} field of view and the pipeline apertures are shown in Fig.~\ref{fig:TESS_apertures} in Appendix~\ref{appendix:TESS_apertures}. This aperture ($\approx21\arcsec\times21\arcsec$) covers the positions of TOI-1710 and other stars. 
Even though the other stars are at least $4\:\mathrm{mag}$ fainter than TOI-1710 in the \textit{TESS} band, their flux contribution is non-negligible as presented in Sect.~\ref{sec:additional_photometry}, ground-based observations were conducted in order to ensure that the transit does not originate from the background stars.

Four similar planetary-transit features were observed on 18 December 2019, 12 January and 30 June 2020, and 29 June 2021; on 2 August 2020 TOI-1710.01 was identified as a promising candidate for a transiting planet with a period of $24.28305\,\pm\,0.00015$~days, a duration of $\sim$4h, a transit depth of $2.52\text{\textperthousand}\,\pm\,0.15\text{\textperthousand}$, and a radius in stellar radii of $5.02\%\,\pm\,0.15\%$, corresponding to an estimated radius of $5.35\,\pm\,0.30~{\rm R}_{\rm \earth}$. 
The candidate passed all the tests from the Threshold Crossing Event (TCE) Data Validation Report \citep[DVR;][]{2018PASP..130f4502T,2019PASP..131b4506L}, including the difference image centroiding test, which located the source of the transit signal to within $0.74\pm2.5~\mathrm{arcsec}$ of the target star.

\subsection{Radial velocities}
\label{sec:RV_follow_up}

\subsubsection{Reconnaissance spectroscopy with TRES}
\label{sec:recon_spectroscopy_TRESS}

On 24 February 2020 and 6 March 2020, we collected two reconnaissance spectra taken with the Tillinghast Reflector Échelle Spectrograph (TRES), a fiber-fed (sky+object+calibration spectrum) optical échelle spectrograph with a resolving power R of 44\,000 installed on the 1.5-meter Tillinghast telescope at the Smithsonian Astrophysical Observatory's Fred~L.~Whipple Observatory (FLWO) on Mt.~Hopkins in Arizona, USA. The spectra were extracted as described in \citet{2010ApJ...720.1118B} and the spectra were cross-correlated against each other, order by order, using the strongest observation as a template. 
The two TRES observations were obtained near opposite quadratures in the same orbital cycle.
The multi-order analysis of the spectra shows a 2~m/s difference between the two spectra, well within the uncertainty on the measurement (27~m/s), contributing to eliminating stellar companions, brown dwarfs, and massive planets as the source of the transits. 
These last were confirmed with the larger phase coverage from SOPHIE and HARPS-N (see below). The TRES spectra were visually inspected to ensure that the spectra did not show any sign of light from another star causing a double-lined spectrum. The corresponding cross-correlation functions (CCFs) were also inspected and no asymmetries were found.

\begin{figure*}
    \centering
    \includegraphics[trim=0.20cm 0.cm 0.0cm 0.cm, clip=true, scale=0.70]{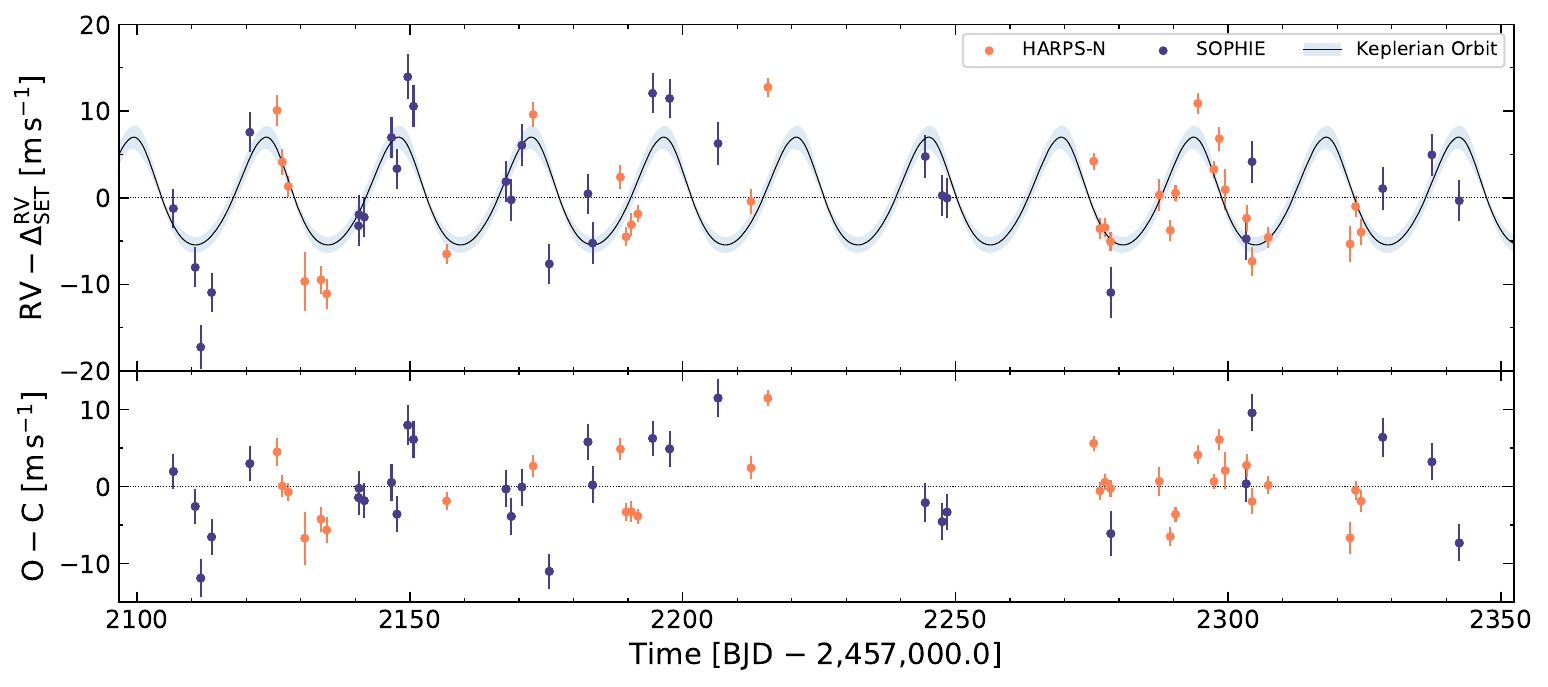}
    \caption{RV evolution of TOI-1710. \textit{Top panel:} Long-term variation in the RV of TOI-1710 determined from the SOPHIE (in blue) and the HARPS-N (in orange) observations relative to a fitted instrumental offset (see text for details). The black solid line and its light blue overlay show the best-fit RV model presented in this study (Sect.~\ref{sec:transit_keplerian_characterization}) with its relative 1$\upsigma$ uncertainty, respectively. \textit{Bottom panel:} RV residuals.}\label{fig:RV}
    \vspace{10 pt}
\end{figure*}

\subsubsection{Spectroscopic follow-up with SOPHIE}
\label{sec:RV_follow_up_SOPHIE}

After its identification from \textit{TESS} photometry, we launched a RV follow-up of TOI-1710 with the SOPHIE (Spectrographe pour l'Observation des Phénomènes des Intérieurs stellaires et des Exoplanètes) high-resolution spectrograph installed at the 1.93-meter telescope of the Observatoire de Haute-Provence, France. It is a stabilized échelle spectrograph fed with a 3\textquotedbl~fiber aperture dedicated to high-precision RV measurements \citep[][]{2008SPIE.7014E..0JP,2009A&A...505..853B,2013A&A...549A..49B}. 
The goal was to establish the presumed planetary nature of the transiting candidate and, in case of a positive detection, to then characterize the planet by measuring, among other parameters, its mass and orbital eccentricity. 
We used the high-resolution (HR) mode of SOPHIE with a resolving power $\mathrm{R}=75\:000$ and fast readout mode, still allowing a clear throughput for this relatively bright star. 

We excluded from our analysis six spectra with a signal-to-noise ratio (S/N) per pixel at 550~nm below 40. The remaining 30 spectra have a homogeneous S/N of $50 \pm 2$, with typical exposure times of 14 minutes (ranging from 8 to 28 minutes depending on weather conditions). Four exposures were significantly polluted by sky moonlight and corrected using the second SOPHIE aperture placed on the sky background following the method described by \citet{2010A&A...520A..65B}, among others. We used the SOPHIE pipeline \citep{2009A&A...505..853B} to extract the spectra from the detector images, cross-correlate them with a G2-type numerical mask (the most similar to the spectral type of TOI-1710 among the available templates), which produced clear CCFs, then fit the CCFs by Gaussians to derive the radial velocities \citep{1996A&AS..119..373B,2002A&A...388..632P}. We corrected them for the RV constant master which takes small additional instrumental drifts into account, following \citet{2015A&A...581A..38C}. That correction has a dispersion of 1.5 m/s over the time span of our observations and does not significantly change our measurements. The bisector inverse slope (BIS) of each CCF was also computed following \citet{2001A&A...379..279Q}. All 39 SOPHIE spectral orders (covering the wavelength range 3872\:\AA\ -- 6943\:\AA) were used for the cross-correlations.

Our final SOPHIE data set holds 30 RVs, acquired between 14 September 2020 and 7 May 2021, with precisions ranging from $2.3$ to $2.9~\mathrm{m\,s^{-1}}$ depending on the S/N, with a median value of $2.4~\mathrm{m\,s^{-1}}$, while the RMS of the RVs is $7.4~\mathrm{m\,s^{-1}}$. 
These points are reported with their corresponding uncertainties, full width at half maximum (FWHM), BIS, spectral index $\log R^{\prime}_{\rm HK}$, and S/N in Appendix~\ref{appendix:RVs}, Table~\ref{tab:SOPHIE_RV}. The RV time series is displayed in the upper panels of Fig.~\ref{fig:RV} and Fig.~\ref{fig:RV_phase} in blue.

\subsubsection{Spectroscopic follow-up with HARPS-N}
\label{sec:RV_follow_up_HARPSN}

In addition, we collected 31 high-resolution spectra of TOI-1710, between 3 October 2020 and 19 April 2021, using the High Accuracy HARPS-N (Radial Velocity Searcher for the Northern Hemisphere) spectrograph \citep{2012SPIE.8446E..1VC, 2014SPIE.9147E..8CC} mounted at the Italian Telescopio Nazionale Galileo 3.58-meter telescope located at the Roque de los Muchachos Observatory on the island of La Palma, Canary Islands, Spain. 
The high-resolution ($\mathrm{R}=115\:000$) échelle spectrograph HARPS-N is stabilized and well controlled in pressure and temperature to minimize instrumental drifts; it can reach sub-$\mathrm{m\,s^{-1}}$ RV precision \citep{2018A&A...620A..47D}. 

The observations of TOI-1710 were carried out within the framework of the Italian project Global Architecture of Planetary Systems (GAPS) (e.g., \citealt{2016frap.confE..69B}). 
The instrument was used with exposure times of 15 minutes and a 1\textquotedbl~fiber on the star and another one placed to monitor the sky-background. 
The spectra have a median S/N per pixel of $69$, measured at the échelle order with mean wavelength $\sim$5700$\:\text{\AA}$. 
The radial velocities of TOI-1710 were extracted using \textsc{v3.7} of the Data Reduction Software (DRS) pipeline, which is based on the CCF method. 
As for SOPHIE, each HARPS-N spectrum was correlated with a mask optimized for a star with spectral type G2V. 
We ran the offline version of the DRS, which is implemented at the INAF Trieste Observatory\footnote{\texttt{\href{http://ia2.inaf.it}{http://ia2.inaf.it}}}, through the YABI workflow interface \citep{YABI}.

Our final HARPS-N data set has 31 RVs with a precision ranging from $0.9$ to $3.4~\mathrm{m\,s^{-1}}$ depending on S/N, with a median value of $1.4~\mathrm{m\,s^{-1}}$, while the RMS of the RVs is $6.1~\mathrm{m\,s^{-1}}$. 
These points are reported with their corresponding uncertainties, FWHM, BIS, spectral indices $I_{\rm Ca\:\sc{ii}}$ and $I_{\rm H\upalpha}$, and S/N in Appendix~\ref{appendix:RVs}, Table~\ref{tab:HARPSN_RV}, and displayed in the upper panels of Fig.~\ref{fig:RV} and Fig.~\ref{fig:RV_phase} in orange.

\subsection{Additional photometric measurements}
\label{sec:additional_photometry}

We observed two photometric transits from the ground-based telescopes George Mason University Observatory (GMU) and Las Cumbres Observatory Global Telescope (LCOGT). They confirmed the \textit{TESS} transit detections and periods. Thanks to their angular resolution, which is better than that of \textit{TESS}, they also confirmed that TOI-1710 is the transited object.

\begin{figure}[ht]
\centering
\includegraphics[trim=0.00cm 0.cm 0.0cm 0.cm, clip=true, scale=0.49]{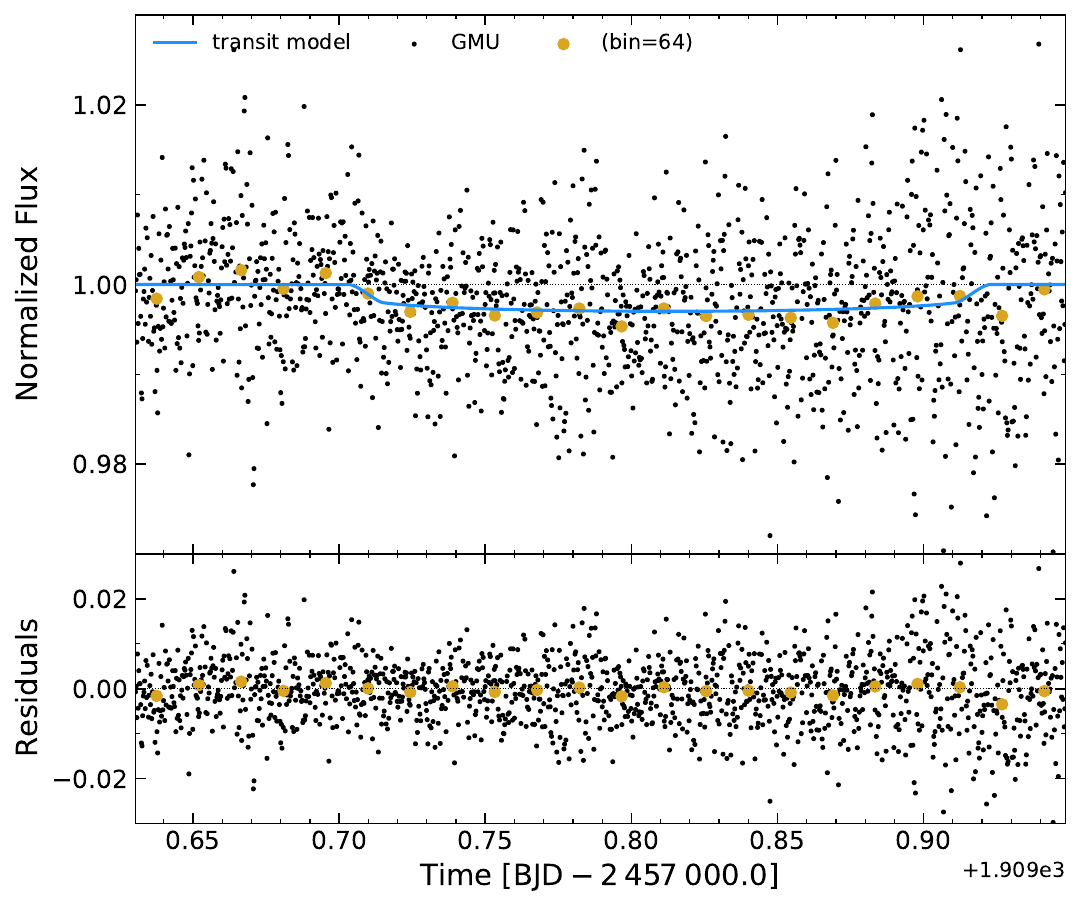}
\caption{Transit LC of TOI-1710 collected by GMU (upper panel). The black dots represent the airmass-detrended photometric data points, and the larger colored dots the corresponding binned LC. The blue line shows the transit profile of the best-fit model presented in this study (Sect.~\ref{sec:transit_keplerian_characterization}) and is used to compute the residuals presented in the bottom panel.}\label{fig:LC_GMU}
\end{figure}

\begin{figure}[ht]
\centering
\includegraphics[trim=0.00cm 0.cm 0.0cm 0.cm, clip=true, scale=0.48]{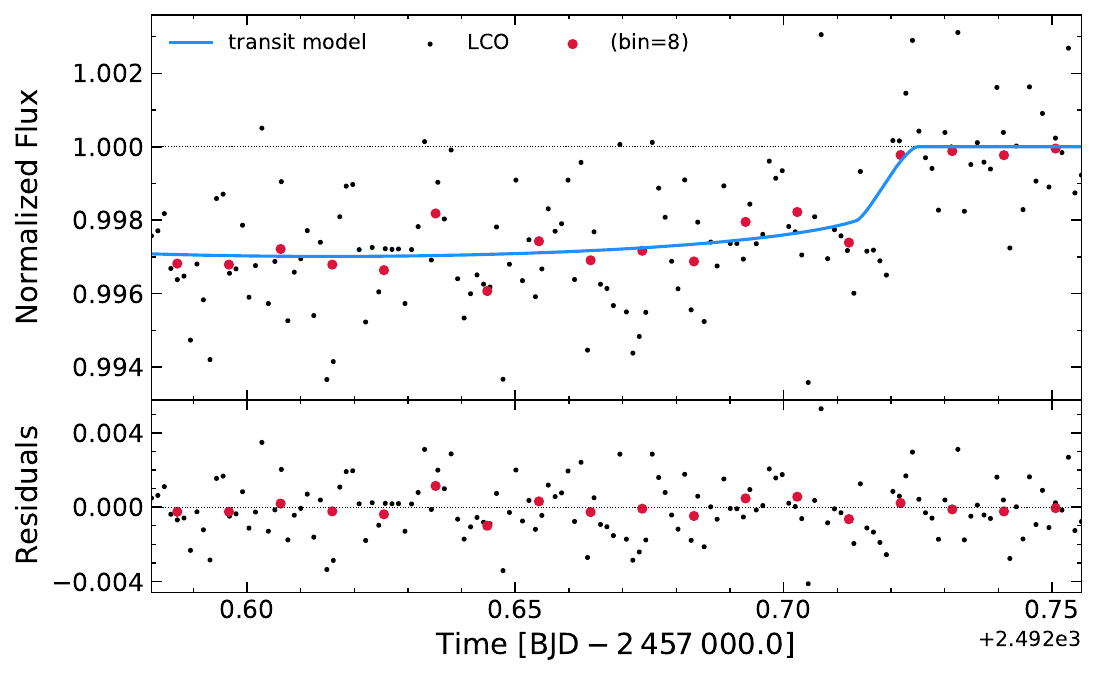}
\caption{Transit LC of TOI-1710 collected by LCOGT (upper panel). The black dots represent the airmass-detrended photometric data points, and the larger colored dots the corresponding binned LC. The blue line shows the transit profile of the best-fit model presented in this study (Sect.~\ref{sec:transit_keplerian_characterization}) and is used to compute the residuals presented in the bottom panel.}\label{fig:LC_LCO}
\end{figure}

\subsubsection{GMU 0.8 m}

Within the scope of the \textit{TESS} Follow-up Observing Program (TFOP) Subgroup 1 (SG1), we observed a full transit of TOI-1710 b in R band on 29 February 2020 from the GMU 0.8 m telescope near Fairfax, USA (Fig.~\ref{fig:LC_GMU}). The telescope is equipped with a $4096\times4096$ SBIG-16803 camera having an image scale of $0\farcs35$ per pixel, resulting in a $23\arcmin\times23\arcmin$ field of view. The images were calibrated and photometric data were extracted using {\tt AstroImageJ} \citep{Collins:2017}. The images have typical stellar point spread functions with a FWHM of $\sim4\arcsec$, and circular photometric apertures with radius $5\farcs3$ were used to extract the differential photometry. The photometric apertures exclude most of the flux from the nearest Gaia EDR3 neighbor (TIC\:705239235) $8\farcs8$ away. Although the timing of egress is not precise due to higher airmass at the end of observations, a $\sim$2800~ppm ingress was detected on target.

\subsubsection{LCOGT 1 m}

Also within the scope of the TFOP SG1, we observed a partial transit of TOI-1710 b in Pan-STARRS $y$ band ($\lambda_{\rm eff} = 9613$\,\AA, ${\rm FWHM} =628$\,\AA) on 5 October 2021 from the LCOGT \citep[][]{Brown:2013} 1.0 m network node at Teide Observatory (Fig.~\ref{fig:LC_LCO}). The 1 m telescopes are equipped with $4096\times4096$ SINISTRO cameras having an image scale of $0\farcs389$ per pixel, resulting in a $26\arcmin\times26\arcmin$ field of view. The images were calibrated by the standard LCOGT {\tt BANZAI} pipeline \citep{McCully:2018}, and photometric data were extracted using {\tt AstroImageJ} \citep{Collins:2017}. The images were focused and have mean stellar point spread functions with a FWHM of $2\farcs3$, and circular photometric apertures with radius $5\farcs4$ were used to extract the differential photometry. The photometric apertures exclude most of the flux from the nearest Gaia EDR3 neighbor (TIC\:705239235) $8\farcs8$ away. A $\sim$2900~ppm egress was detected on target.


\section{Host star fundamental properties}
\label{sec:star_prop}

TOI-1710 is a star with an apparent magnitude $V$ of $9.62$. It is located at a distance of $81.1\pm0.1~\mathrm{pc}$, as obtained via the updated parallax from the Gaia mission early data release~3 (Gaia EDR3, \citet{2016A&A...595A...1G,2021A&A...649A...1G}). The stellar properties are listed in Table~\ref{tab:star}. For the stellar analysis detailed in this section, as for the rest of this study, we used the values of the units ${\rm R}_{\rm \sun}$, ${\rm M}_{\rm \sun}$, ${\rm R}_{\rm \earth}$, and ${\rm M}_{\rm \earth}$ from the \texttt{astropy}\footnote{\texttt{\href{https://docs.astropy.org/en/stable/constants/index.html}{https://docs.astropy.org/en/stable/constants}}}~v5.0 Python package.

\begin{table}
\newcolumntype{C}{ @{}>{{}}c<{{}}@{} }
\centering
\caption[]{Stellar parameters of the planet-hosting star TOI-1710.}\label{tab:star}
\begin{tabular}{llrClc}
    \toprule
    Parameter & &  \multicolumn{3}{c}{Value (±1$\upsigma$)} & Source \\ 
    \hline
    
    \multicolumn{2}{l}{\footnotesize{\underline{Identifying Information}}} & & \rule{0pt}{2.6ex} \\ 
    TOI &  & \multicolumn{3}{c}{TOI-1710} & \textit{TESS} \\
    TIC ID &  & \multicolumn{3}{c}{445805961} & \textit{TESS} \\
    Tycho &  & \multicolumn{3}{c}{4525-1009-1} & Tycho \\
    2MASS ID & \multicolumn{4}{r}{J06170789+7612387} & 2MASS \\
    Gaia ID & \multicolumn{4}{r}{1116613156757308928} & Gaia DR1 \\
    Gaia ID & \multicolumn{4}{r}{1116613161053977472} & Gaia DR2 \\
    Gaia ID & \multicolumn{4}{r}{1116613161053977472} & Gaia \tiny{EDR3} \\
 
    \multicolumn{2}{l}{\footnotesize{\underline{Astrometric Parameters}}} \rule{0pt}{2.6ex} \\ 
    R.A. & [J2000]    & \multicolumn{3}{c}{06:17:07.862} & Gaia DR2 \\
    Dec & [J2000] & \multicolumn{3}{c}{+76 12 38.810} & Gaia DR2 \\
    Parallax & $\mathrm{[mas]}$ & $12.3247$ & $\:\pm\:$ & $0.0103$     & Gaia \tiny{EDR3} \\
    Distance & $\mathrm{[pc]}$ & $81.1376$ & $\:\pm\:$ & $0.0675$ & Gaia \tiny{EDR3} \\
    
    \multicolumn{2}{l}{\footnotesize{\underline{Magnitudes}}} \rule{0pt}{2.6ex} \\ 
    $B_{T}$ &     & $10.35$ & $\:\pm\:$ & $0.03$ & Tycho \\ 
    $B$ (Johnson) &     & $10.15$ & $\:\pm\:$ & $0.03$ & APASS \\
    $V$ &     & $9.62$ & $\:\pm\:$ & $0.02$ & Tycho \\ 
    $G$   &     & $9.3598$ & $\:\pm\:$ & $0.0002$ & Gaia DR2 \\
    $J$   &     & $8.319$ & $\:\pm\:$ & $0.019$ & 2MASS \\
    $H$   &     & $8.003$ & $\:\pm\:$ & $0.034$ & 2MASS \\
    $K_{\rm S}$ &     & $7.959$ & $\:\pm\:$ & $0.026$ & 2MASS \\
    $W1$  &     & $7.883$ & $\:\pm\:$ & $0.024$ & AllWISE \\
    $W2$  &     & $7.938$ & $\:\pm\:$ & $0.020$ & AllWISE \\
    $W3$  &     & $7.915$ & $\:\pm\:$ & $0.020$ & AllWISE \\
    $W4$  &     & $7.997$ & $\:\pm\:$ & $0.215$ & AllWISE \\

    \multicolumn{2}{l}{\footnotesize{\underline{Bulk Parameters}}} \rule{0pt}{2.6ex} \\ 
    Luminosity $L$ & $[{\rm L}_{\rm \sun}]$ & \multicolumn{3}{c}{$0.895$}  & Gaia DR2 \\
    Mass $M_{\rm \star}$ & $[{\rm M}_{\rm \sun}]$ & \multicolumn{3}{c}{$0.984^{+0.050}_{-0.059}$} & Sect.\:\ref{sec:star_prop} \\
    Radius $R_{\rm \star}$ & $[{\rm R}_{\rm \sun}]$ & \multicolumn{3}{c}{$0.968^{+0.016}_{-0.014}$} &  Sect.\:\ref{sec:star_prop} \\
    $T_{\rm eff}$ & $[\mathrm{K}]$ & $5665$ & $\:\pm\:$ & $55$ & Sect.\:\ref{sec:star_prop} \\
    $\log{g_{\rm \star}}$ & $[\mathrm{cgs}]$ & $4.46$ & $\:\pm\:$ & $0.10$ & Sect.\:\ref{sec:star_prop} \\
    $\xi$ & $[\mathrm{km\,s^{-1}}]$ & $0.86$ & $\:\pm\:$ & $0.10$ & Sect.\:\ref{sec:star_prop} \\
    $\mathrm{[Fe/H]}$ & $\mathrm{[dex]}$ & $0.10$ & $\:\pm\:$ & $0.07$ & Sect.\:\ref{sec:star_prop} \\
    Spectral type$^{(*)}$ &  & \multicolumn{3}{c}{G5V} & Sect.\:\ref{sec:star_prop} \\
    $\rho_{\rm \star}$ & $[\mathrm{g\,cm^{-3}}]$ & \multicolumn{3}{c}{$1.53^{+0.15}_{-0.16}$} & Sect.\:\ref{sec:star_prop} \\
    $v \sin i_{\rm \star}$ & $[\mathrm{km\,s^{-1}}]$ & $2.3$ & $\:\pm\:$ & $0.4$ & Sect.\:\ref{sec:star_prop} \\
    $P_{\rm rot}$ & $[\mathrm{d}]$ & $22.5$ & $\:\pm\:$ & $2.0$ & Sect.\:\ref{sec:star_prop} \\
    $\log R^\prime_{\rm HK}$ & $\mathrm{[dex]}$ &  $-4.78$ & $\:\pm\:$ & $0.03$ & Sect.\:\ref{sec:star_prop} \\
    Age $\tau_{\rm \star}$ & $[\mathrm{Gyr}]$ & \multicolumn{3}{c}{$4.2^{+4.1}_{-2.7}$}  & Sect.\:\ref{sec:star_prop}\rule[-1.2ex]{0pt}{0pt} \\
    $u_{\rm \star}$ &  & $0.6313$ & $\:\pm\:$ & $0.0027$ & Sect.\:\ref{sec:star_prop}\rule[-1.2ex]{0pt}{0pt} \\
    $v_{\rm \star}$ &  & $0.0827$ & $\:\pm\:$ & $0.0025$ & Sect.\:\ref{sec:star_prop}\rule[-1.2ex]{0pt}{0pt} \\

    \bottomrule
\end{tabular}
\tablebib{Tycho \citep{2000A&A...355L..27H}; 2MASS \citep{2006AJ....131.1163S}; Gaia Collaboration (\citeyear{2016A&A...595A...1G}, \citeyear{2018A&A...616A...1G}, \citeyear{2021A&A...649A...1G}); APASS \citep{henden2015}; AllWISE \citep{allwise2013}.
}
\tablefoot{$^{(*)}$~Spectral type defined according to the stellar spectral classification of \citet{2009ssc..book.....G}.}
\end{table}

\subsection{Spectral characterization from SOPHIE spectra}
\label{sec:spectral_characterization}

We analyzed the co-added individual SOPHIE spectrum of the planet-hosting star TOI-1710, with S/N of $\sim$100 per pixel at $670~\mathrm{nm}$ in the continuum (near the \ion{Li}{i} line region). 
We performed the spectral analysis using the 2019 version of the \texttt{MOOG}\footnote{\texttt{\href{https://www.as.utexas.edu/~chris/moog}{https://www.as.utexas.edu/\char`\~chris/moog}}} package \citep{1973ApJ...184..839S}. 
Following the methodology described in detail by \citet{2013A&A...556A.150S}, \citet{2021arXiv210904781S}, and references therein, the stellar atmospheric parameters and respective uncertainties were derived from the equivalent widths (EW) of 246+31 Fe~{\sc i} and Fe~{\sc ii} weak lines carefully selected, and assuming ionization and excitation equilibrium. The EWs were measured with \texttt{ARES}\footnote{\texttt{\href{http://www.astro.up.pt/~sousasag/ares}{http://www.astro.up.pt/\char`\~sousasag/ares}}} \citep{2015A&A...577A..67S}. This procedure leads to the following stellar parameters: 
effective temperature $T_{\rm eff}=5693\pm66~\mathrm{K}$, surface gravity $\log{g_{\rm \star}}=4.40\pm0.12$ (cgs), and iron abundance $\text{[Fe/H]}=0.11\pm0.06~\mathrm{dex}$. 
This is in agreement with the astrometric $\log{g_{\rm \star}}$ derived making use of the Gaia parallax and luminosity \citep[see][]{2004A&A...415.1153S}, which gives a value of $4.50~\mathrm{dex}$. 

Using the \citet{2010A&ARv..18...67T} calibration leads to a stellar mass $M_{\rm \star}$ of $1.00\pm0.05~{\rm M}_{\rm \sun}$ and a stellar radius $R_{\rm \star}$ of $1.10\pm0.17~{\rm R}_{\rm \sun}$, thus corresponding to an average density $\rho_{\rm \star}$ of $1.06~\mathrm{g\,cm^{-3}}\lesssim\:\rho_{\rm \sun}$. We reevaluate these fundamental stellar parameters via other methods in Sect.~\ref{sec:analysis_sed} and \ref{sec:models_parametrization_without_GPs}.
From the value of its metallicity and comparison of the location of TOI-1710 in the Hertzsprung–Russell diagram with evolutionary tracks, we can ascertain its spectral type to be G5V, still lying on the main sequence according to the stellar spectral classification of \citet{2009ssc..book.....G}. 

The rotation velocity was deduced from the CCF width as $v\sin i_{\rm \star}=3.2\pm1.0\:\mathrm{km\,s^{-1}}$ \citep[for the method applied here, see][]{2010A&A...523A..88B}. 
Similarly, by studying the $\ion{Ca}{ii}~\rm{H\,\&\,K}$ Fraunhofer core lines (centered at 3968.5 and 3933.7~\text{\normalfont \AA}) in the SOPHIE spectra, we can measure the value of the activity tracer $\log R^{\prime}_{\rm HK}=-4.8\pm0.1~\mathrm{dex}$ \citep[see][]{2010A&A...523A..88B}, indicating the possible contribution of stellar activity to this spectroscopic data.

\subsection{Fundamental parameters derived from HARPS-N spectra and spectral energy distribution}
\label{sec:analysis_sed}

\begin{figure}[htbp]
    \centering
    \includegraphics[trim=0.00cm 0.00cm 0.00cm 0.00cm, clip=true, width=0.49\textwidth]{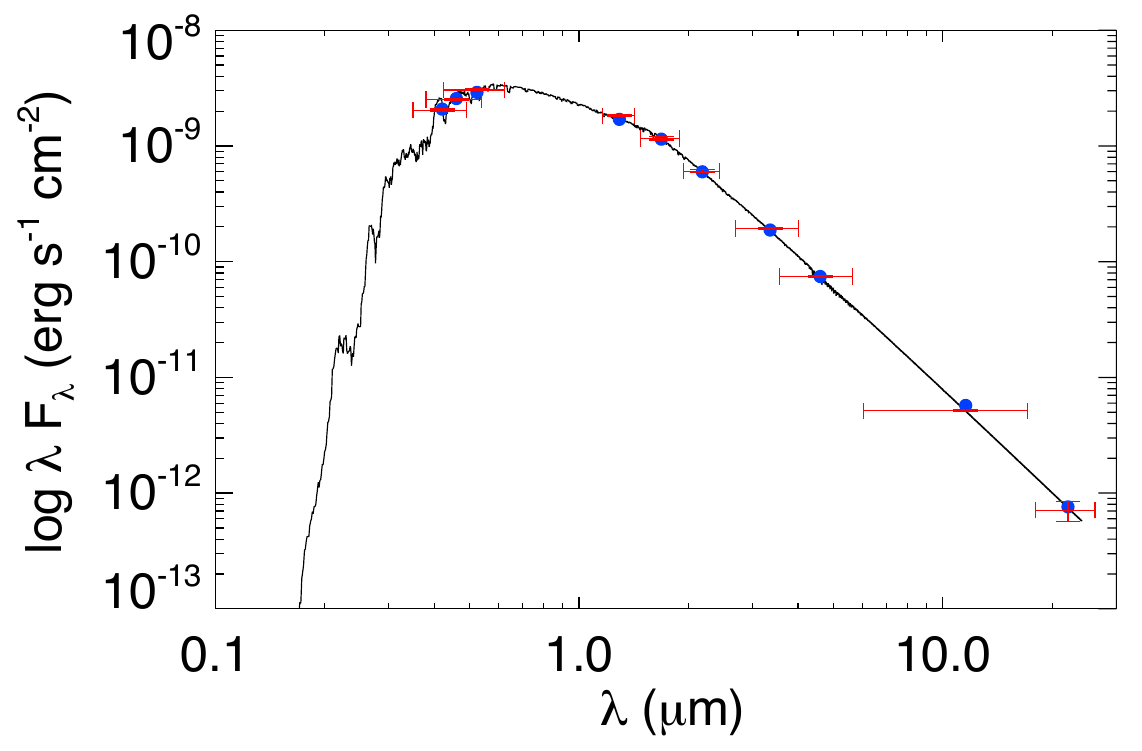}
    \caption{Spectral energy distribution of TOI-1710. The red markers represent the photometric measurements; the vertical error bars correspond to the reported measurement uncertainties from the catalog photometry (Table~\ref{tab:star}, Magnitudes). The horizontal error bars depict the effective width of each passband. The black curve corresponds to the most likely stellar atmosphere model. Blue circles identify the model fluxes over each passband.}
    \label{fig:SED_fit}
\end{figure}

From the HARPS-N spectra, we derived $T_{\rm eff}$, $\log{g_{\rm \star}}$, [Fe/H], and the microturbulence velocity ($\xi$) through a standard method based on the EWs of iron lines measured on coadded HARPS-N spectra.
We then used the \citet{castellikurucz2003} grid of model atmospheres and the spectral analysis package \texttt{MOOG} (2017 version; \citealt{sneden1973}). The 
$T_{\rm eff}$ was derived by imposing that the abundance of \ion{Fe}{i} is not dependent on the line excitation potentials, $\xi$ by obtaining the independence between the \ion{Fe}{i} abundance and EWs, and $\log{g}$ by the \ion{Fe}{i}/\ion{Fe}{ii} ionization equilibrium condition. 
With this method we determined the following stellar atmospheric parameters: $T_{\rm eff}=5665\pm55$ K, $\log{g}=4.46\pm0.10$, $\xi=0.86\pm0.10~\mathrm{km\:s^{-1}}$, and [Fe/H]$=0.10\pm0.07~\mathrm{dex}$. 
Employing a similar method to that used on the SOPHIE spectra, we measure a value for $\log R^{\prime}_{\rm HK}$ of $-4.78\pm0.03~\mathrm{dex}$. 
The $v\sin i_{\rm \star}$ from HARPS-N data was measured with the same
\texttt{MOOG} code and applying the spectral synthesis method after fixing the macroturbulence to the value of 2.9~km/s from the relationship by \citet{2014MNRAS.444.3592D}. We find a 
projected rotational velocity of $2.3\pm0.4~\mathrm{km/s}$.
These values are consistent with the results obtained from the SOPHIE spectra exposed in Sect.~\ref{sec:spectral_characterization} and confirm that TOI-1710 is not a rapid rotator. 
Nevertheless, we adopted the atmospheric parameters derived from the HARPS-N spectra, which have a higher resolution and a higher S/N than the SOPHIE spectra. 
We use these values to perform the analysis of the spectral energy distribution (SED) next. 

To determine the radius, mass, and age of TOI-1710, we fitted the stellar SED and used the \texttt{MESA}\footnote{\texttt{\href{http://mesa.sourceforge.net}{http://mesa.sourceforge.net}}} Isochrones \& Stellar Tracks (MIST) \citep{2016ApJS..222....8D} through the \texttt{ExofastV2}\footnote{\texttt{\href{https://github.com/jdeast/EXOFASTv2}{https://github.com/jdeast/EXOFASTv2}}} code \citep{2017ascl.soft10003E, Eastman2019} by imposing Gaussian priors on the spectroscopically derived $T_{\rm eff}$ and [Fe/H], and on the stellar parallax from the Gaia EDR3 value $\pi=12.3247\pm0.0103$~mas \citep{2018A&A...616A...1G}. 
Using the value of the parallax from the Gaia DR2 does not significantly change the results.
For the SED we considered the Johnson $B$ magnitude, the Tycho $B_{\rm T}$ and $V$ magnitudes, the 2MASS $J$, $H$, and $K$ magnitudes, and the WISE magnitudes, corresponding to a wavelength range $0.4\text{--}30~\mathrm{\upmu m}$ (see Table~\ref{tab:star} and Fig.~\ref{fig:SED_fit}). 
We found $R_\star=0.968^{+0.016}_{-0.014}~R_\odot$, $M_\star=0.984^{+0.050}_{-0.059}~M_\odot$, and a poorly constrained stellar age of $4.2^{+4.1}_{-2.7}$~Gyr. 
These values are in agreement within 2$\upsigma$ of the ones derived in Sect.~\ref{sec:spectral_characterization}. We also ran a stellar parameter multi-order analysis from the TRES spectra (Sect.~\ref{sec:recon_spectroscopy_TRESS}) using the Stellar Parameter Classification (SPC) tool introduced by \citet{2012Natur.486..375B} and obtained the following agreeing values: $T_{\rm eff}=5705\pm50~\mathrm{K}$, $\log{g_{\rm \star}}=4.48\pm0.10$, $\mathrm{[m/H]}=0.04\pm0.08$, $v\sin i_{\rm \star}=3.1\pm0.5~\mathrm{km/s}$. 
We chose for this study to retain the stellar properties resulting from the analysis of the HARPS-N spectra because they are more accurately defined. These values are then used as priors in the global fit with \texttt{celerite2} and \texttt{PyMC3} in Sect.~\ref{sec:transit_keplerian_characterization} to derive the fundamental parameters of the planet and its orbit (mass, radius, semi-major axis). All the adopted stellar parameters are listed in Table~\ref{tab:star}. 

\subsection{Limb-darkening profile}
\label{sec:limb_darkening_profile}

Supplementing this spectral analysis, we determined the limb-darkening profile of TOI-1710 via the Python Limb Darkening
Toolkit package \citep[\texttt{PyLDTk};][]{2015MNRAS.453.3821P} for calculating stellar limb-darkening profiles and model-specific coefficients for arbitrary passbands using the stellar spectrum model library PHOENIX by \citet{2013A&A...553A...6H}. According to this method, and following the two-parameter limb-darkening parameterization of \citet{2013MNRAS.435.2152K}, the stellar parameters presented above lead to the limb-darkening quadratic law coefficients $(u_{\rm \star},\:v_{\rm \star})=(0.6313,\:0.0027)\pm(0.0827,\:0.0025)$.


\subsection{Activity and rotation}
\label{sec:rotational_period_age}

\begin{figure*}[htbp]
    \centering
    \includegraphics[trim=0.0cm 0.0cm 0.0cm 0.0cm, clip=true, width=\textwidth]{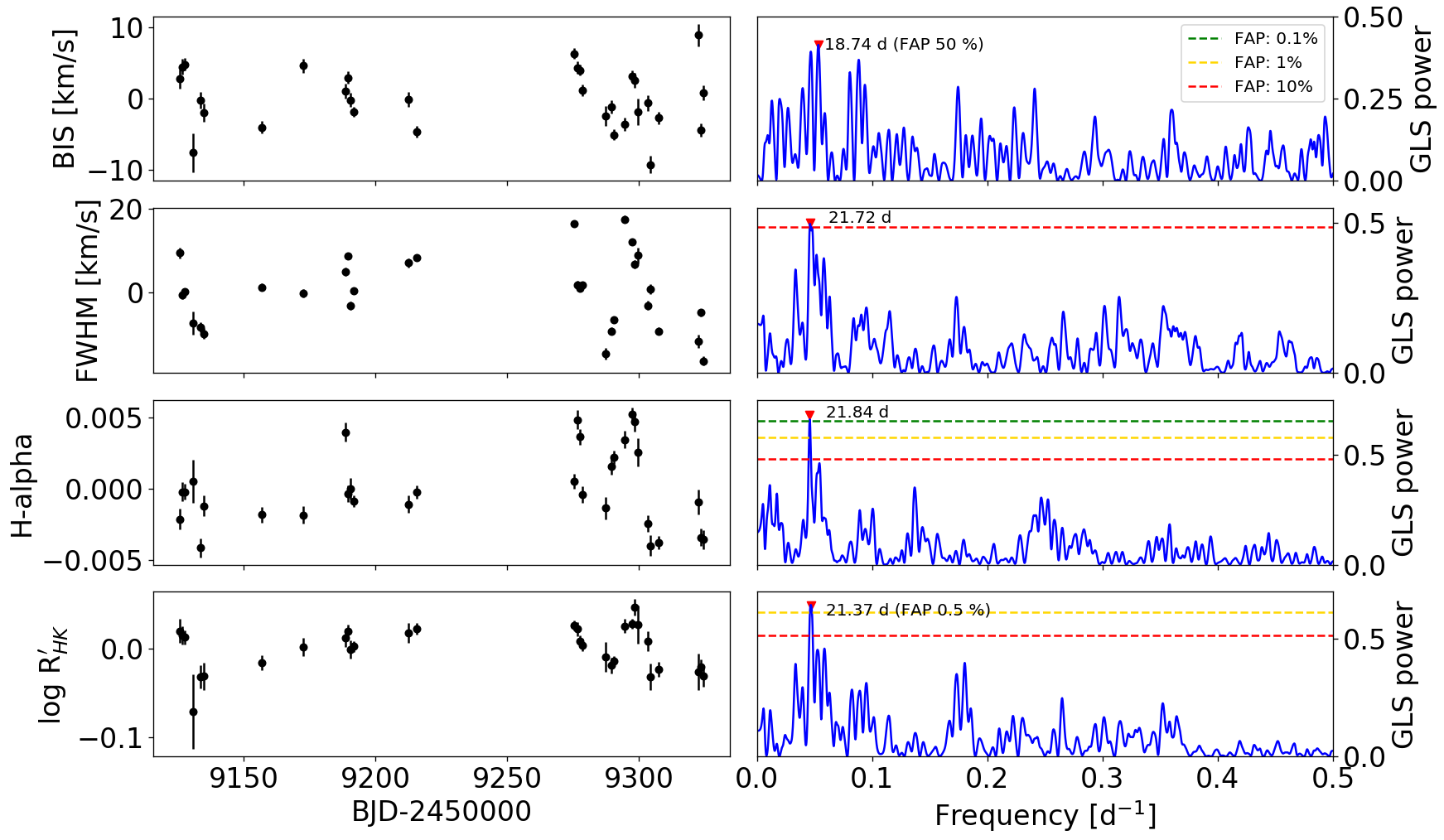}
    \caption{Spectroscopic activity diagnostics derived from HARPS-N spectra of TOI-1710. \textit{Left panels:} Time series of the spectroscopic activity diagnostics. The average value was subtracted from the original data. \textit{Right panels:} Generalized Lomb--Scargle periodograms of each time series, with the period corresponding to the highest power peak indicated. The levels of the false alarm probabilities, derived through a bootstrap analysis, are indicated as horizontal dashed lines.}
    \label{fig:GLS_periodogram_act_ind}
\end{figure*}

The \textit{TESS} LC of TOI-1710 (sectors 19, 20, 26, and 40 together), as shown in the upper panel of Fig.~\ref{fig:LC}, does not show a clear periodic or quasi-periodic variability. 
The photometric data used in this study are taken from the PDC-SAP flux, namely the SAP flux from which long-term trends were removed using co-trending basis vectors; they are therefore not optimal when searching for long-term variations such as activity-related variations.
Even so, the observed out-of-transit modulations appear different in the different sectors, and show sub-mmag variations. We tried to estimate the rotational period from photometry using two methods, after masking the transits of TOI-1710\,b and averaging the data in bins of four hours to speed up the analysis, without any loss of information. We first applied a Gaussian process (GP) quasi-periodic regression, but the analysis did not provide evidence for a well-defined rotation period. Then we decided to perform a harmonic analysis by fitting the LC with a Fourier series using the software \textsc{Period04} \citep{Lenz05}. We analysed sectors 19+20, 26, and 40 separately, taking into account the changes in the LC behavior due to the large time gaps between the sectors, and we fitted up to three sinusoids. We found that the dominant frequencies for each of group of data are different: [0.0873, 0.109, 0.148]~$\mathrm{d^{-1}}$ (P=11.45~d; 9.17~d; 6.76~d); [0.183, 0.151, 0.117]~$\mathrm{d^{-1}}$ (P=5.46~d; 6.62~d; 8.55~d); and [0.117, 0.291, 0.218]~$\mathrm{d^{-1}}$ (P=8.55~d; 3.44~d; 4.59~d), respectively. Their physical interpretation in terms of stellar rotation is not obvious; therefore, we conclude that \textit{TESS} photometry does not allow us to identify the photometric stellar rotation period.

Alternatively, it is possible to infer the rotational period of the system using the empirical activity-rotation relationship of \citet{2008ApJ...687.1264M}. Knowing the average value of the Ca~{\sc ii}$~\rm{H\,\&\,K}$ chromospheric activity indicator $\log R^{\prime}_{\rm HK}=-4.78\pm0.03$ obtained from the HARPS-N calibration, we derive a stellar rotation period $P_{\rm rot}=20.1\pm2.6~\mathrm{days}$. 

Adopting this rotational period with the empirical rotation-age relations of \citet{2008ApJ...687.1264M} leads to an isochronal age estimate of $\tau_{\rm \star}=3.06\pm0.33~\mathrm{Gyr}$. Because the $v\sin i_{\rm \star}$ is a projected rotational velocity, we note that it is a lower limit on the true rotational velocity, and thus $P_{\rm rot}/\sin i_{\rm \star}$ is an upper limit on the true $P_{\rm rot}$, which implies that the age estimate $\tau_{\rm \star}$ above provides an upper limit to the stellar age of TOI-1710. 

Additionally, we used the generalized Lomb--Scargle (GLS) algorithm \citep{zech09} to determine the frequency content of some commonly used spectroscopic activity diagnostics calculated from the HARPS-N spectra. These are the BIS, FWHM, $\log R^{\prime}_{\rm HK}$ (all calculated by the DRS pipeline), and the index based on the H-alpha chromospheric line, which was derived using version 1.3.7 of the software \texttt{ACTIN} \citep{gomesdasilva18}. The time series and the GLS periodograms of these activity diagnostics are shown in Fig.~\ref{fig:GLS_periodogram_act_ind}. The false alarm probability (FAP) levels up to 10$^{-4}$ were determined through a statistical bootstrap (with replacement) analysis, including 10\,000 permutations of each time series. Except for the BIS, all the other indices have the highest power peak located at 21--22 days, with uncertainties estimated to about two days, and with low FAPs in the case of the H-alpha and $\log R^{\prime}_{\rm HK}$ indices. This result agrees with the expected range derived from the empirical activity-rotation relationship.

Finally, the HARPS-N data from the CCF FWHM and the color index $V-K_{\rm S}=1.66\pm0.05$ leads to a convective turnover time of $\tau_{\rm c}=12.5\pm1.3~\mathrm{days}$ ($\log\tau_{\rm c}=1.10\pm0.01~\mathrm{dex}$), according to the empirical relationship presented by \citet{2011ApJ...743...48W}. Then taking the stellar mass value $M_{\rm \star}$ from Sect.~\ref{sec:analysis_sed} with the color index $B-V=0.73\pm0.05$, we deduce anew the stellar rotation period $P_{\rm rot}=25.5\pm3.8~\mathrm{days}$. 

\begin{figure}[hbtp]
    \centering
    \includegraphics[trim=0.0cm 0.0cm 0.0cm 0.0cm, clip=true, width=0.49\textwidth]{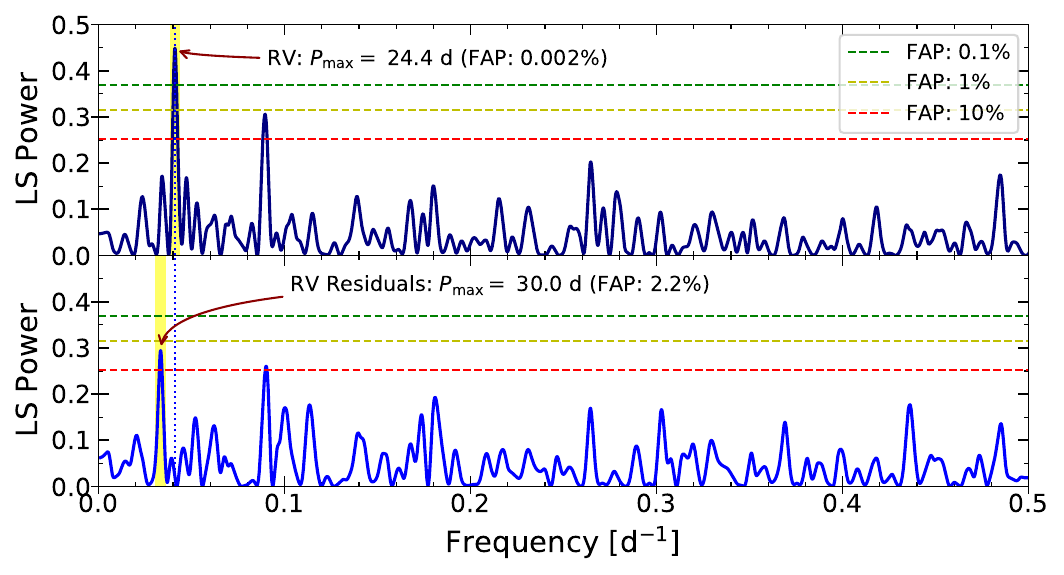}
    \caption{Lomb--Scargle periodograms of the RV (\textit{upper panel}) and RV residuals (\textit{lower panel}) time series (including HARPS-N and SOPHIE data). The frequency corresponding to the period of the transiting planet is indicated by a blue dotted vertical line. The frequency of the maximum peak is highlighted with a thick yellow vertical line. The RV residuals are issued from the best-fit model (see Sect.~\ref{sec:models_parametrization_without_GPs}). The levels of the false alarm probabilities, derived through a bootstrap analysis, are indicated with horizontal dashed lines.}
    \label{fig:GLS_periodogram_RV_res}
\end{figure}

We report in this section how we estimated $P_{\rm rot}$ via three different methods; their respective results are consistent with each other. However, even though these calculations deliver a fair estimate of the proper rotation of the star, each includes systematic effects that are not well understood, so that none of them is an exact determination of it. Thus, to remain conservative, we chose to retain the average value of the three values given above, with the corresponding standard deviation as its uncertainty, namely $\braket{P_{\rm rot}}=22.5\pm2.0~\mathrm{days}$. 

In Fig.~\ref{fig:GLS_periodogram_RV_res}, we show the Lomb--Scargle (LS) periodograms of the HARPS-N + SOPHIE RV and RV residuals time series. The main peak in the upper panel is located at the orbital period of TOI-1710\:b. It is near the stellar rotation period so it is possible to suspect that stellar activity might significantly affect the derived planetary parameters. The two lower panels of Fig.~\ref{fig:GLS_periodogram_RV_res} show the residuals after the best-fit models. In the next section we present the details of these models, and how robust they are with respect to stellar activity.

\section{Characterization of the planetary system}
\label{sec:transit_keplerian_characterization}

For the analysis of the LC and the RV variations of TOI-1710 presented in this section we used together the \textit{TESS} photometric data and the SOPHIE and HARPS-N spectroscopic data for the RV time series. 
Including the GMU and LCO transit LCs to the photometric data does not significantly alter the results; hence, we choose to present the analysis considering the \textit{TESS} LC only.
In this study we used the modeling and fitting method coded in \texttt{exoplanet}\footnote{\texttt{\href{https://docs.exoplanet.codes/en/latest/}{https://docs.exoplanet.codes}}} \citep{2019zndo...1998447F} and \texttt{PyMC3}\footnote{\texttt{\href{https://docs.pymc.io}{https://docs.pymc.io}}} \citep{exoplanet:pymc3}, Python toolkit for Bayesian statistical modeling which focuses on advanced MCMC and variational fitting algorithms.

\subsection{Models and parameterization}
\label{sec:models_parametrization_without_GPs}

The model we developed for the characterization of TOI-1710\:b performs a simultaneous analysis of the RV and LC. Alternative modeling methods and tests are presented in Sect.~\ref{sec:tests_and_validation}. 
This subsection gives the details of the modeling and parameterization chosen and used here to fit the transit profile and the RV variations, the method chosen to determine the prior and posterior distributions, and the results obtained with this procedure. 

\subsubsection{Transit profile}
\label{sec:transit_profile}

As a first step of the analysis of the \textit{TESS} light curve, performing a box least-squares (BLS) periodogram \citep{2002A&A...391..369K} gauges the general parameters of the transit profile \citep{2016ascl.soft07008K}. This procedure is computed here via the functions provided by the \texttt{exoplanet} package. 
The preliminary BLS fit shows one distinct peak leading to an estimate of the orbital period $P$, the time of transit $T_{0}$, the depth $R_{\rm p}^2/R_{\rm \star}^2$ and the duration of the transit which sets the prior values for these parameters. Folding the light curve in phase according to this estimate of $P=24.283429~\mathrm{days}$ and plotting the binned data points allows the profile of the transit to appear.

To complete the description of the transit profile, we additionally introduce in the full LC model the following parameters: the normalized semimajor axis $a/R_{\rm \star}$, the impact parameter $b$, and the quadratic limb-darkening coefficients ($u_{\rm \star}$ and $\:v_{\rm \star}$, Sect.~\ref{sec:limb_darkening_profile}). 
A white noise parameter $\sigma_{\textsf{TESS}}$ is added to the LC model to account for the jitter and instrumental noise.
We then calculated from the parameters the values and errors for the semimajor axis $a$, the planet radius $R_{\rm p}$, and mass $M_{\rm p}$, as well as the mean stellar density $\rho_{\rm \star}$. Indeed, $\rho_{\rm \star}$ can be measured solely from the transit light curve without any measurement of the stellar mass $M_{\rm \star}$ or radius $R_{\rm \star}$ as a result of Kepler’s third law \citep[][]{2003ApJ...585.1038S,2017AJ....154..228S}. 
Since they are defined in the next paragraph in the parameters of the RV model, the eccentricity $e$ and the argument of periastron of the star’s orbit $\omega$ are directly deduced from $\sqrt{e}\,\sin\omega$ and $\sqrt{e}\,\cos\omega$.

\subsubsection{Radial-velocity model}
\label{sec:RV_model}

In order to compensate for an expected offset between SOPHIE and HARPS-N RV measurements, the first element of the analysis of the RV variations is to adjust an independent zero point on each data set. The RV time series becomes $RV\:-\:\braket{RV}_{\textsf{SET}}$.
The parameters describing the eccentric Keplerian orbit are $T_{0}$, $P$, $e$, and $\omega$, as in Sect.~\ref{sec:transit_profile}, and the RV semi-amplitude $K$. 
The work presented by \citet{2018PASP..130d4504F} has shown that reparameterizing the fitted parameters as $T_{0}$, $\ln P$, $\ln K$, $\sqrt{e}\,\sin\omega$, and $\sqrt{e}\,\cos\omega$ forces $P$ and $K>0$, avoids biasing $K$ and prevents the numerical overestimation of $e$, and helps to speed up the MCMC convergence. The photometric epoch $T_{0}$ is set to correspond to the time of the third and most recent transit of the light curve prior to the first RV measurement (i.e., the closest to the RV time series acquired afterward). Two white noise parameters $\sigma_{\textsf{HARPS-N}}$ and $\sigma_{\textsf{SOPHIE}}$ are added to the RV model to account for the jitter and instrumental noise \citep[e.g.,][]{2005blda.book.....G,2009MNRAS.393..969B}.

\subsubsection{Prior distributions and posterior sampling}
\label{sec:prior_distributions_posterior_sampling}

The priors chosen for each of the 13 parameters of the model and parameterization are presented in Appendix~\ref{appendix:prior}, Table~\ref{tab:prior}. For most of the parameters, they are uninformative priors with large bounds. The normal distributions are centered around the values obtained from the BLS periodogram for $T_{0}$, $P$, $R_{\rm p}/R_{\rm \star}$, and $a/R_{\rm \star}$, and around the instrumental respective median values of the signal for $\braket{RV}_{\textsf{HARPS-N}}$ and $\braket{RV}_{\textsf{SOPHIE}}$. Nonrestrictive uniform prior distributions were chosen for the remaining parameters. One exception is the prior on the stellar radius $R_{\rm \star}$ for which we applied the information derived from the spectral analysis in Sect.~\ref{sec:analysis_sed}. We tested the robustness of this prior by releasing the constraint imposed on $R_{\rm \star}$ using the larger distribution $\mathcal{N}(0.968,\:0.064\:{\rm R}_{\rm \sun})$, which did not significantly change our solution.

The posterior distribution was sampled using an MCMC algorithm implemented in \texttt{PyMC3} \citep{exoplanet:pymc3}. We ran the \texttt{PyMC3} algorithm with 16 walkers through 5000 iterations. We discarded the first 1000 steps, considering them as tuning draws. The walkers mixed homogeneously and concentrated before the end of the chains in the same region of the parameter space, around a maximum of the posterior density. 
This indicates that the algorithm has converged. The corner plot (Fig.~\ref{fig:corner}) presents no clear correlations between the parameters of the model, except the expected correlation between $a$ and $R_{\rm p}$, which comes from the uncertainty on parameter $R_{\rm \star}$ and which does not produce additional biases on the results. 

\subsubsection{Results}
\label{sec:results}

\begin{figure}[ht]
    \centering
    \includegraphics[trim=0.25cm 0cm 0cm 0cm, clip=true, width=0.49\textwidth]{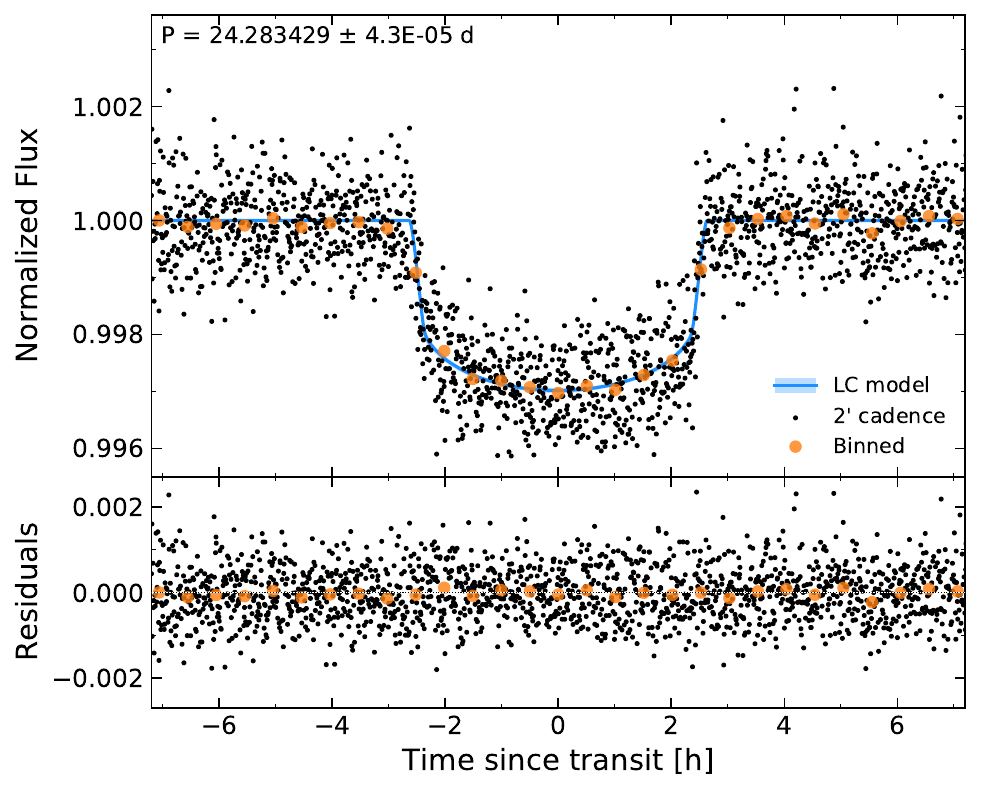}
    \caption{\textit{TESS} light curve of TOI-1710 folded in phase at the best-fit model period $P=24.283429~\mathrm{days}$. Shown are the \textit{TESS} photometric data, acquired with a 2 min cadence (black
dots), and the binned light curve, each bin including 59 data points (i.e., $\sim$2h; larger orange dots). The solid blue line is the best-fit transit model presented in this study (Sect.~\ref{sec:models_parametrization_without_GPs}) and used to compute the residuals in the bottom panel. Its light blue overlay shows its relative 1$\upsigma$ uncertainty (indiscernible from the model on this plot).}\label{fig:LC_phase}
\end{figure}

\begin{figure}[ht]
    \centering
    \includegraphics[trim=0.25cm 0.cm 0.0cm 0.cm, clip=true, width=0.49\textwidth]{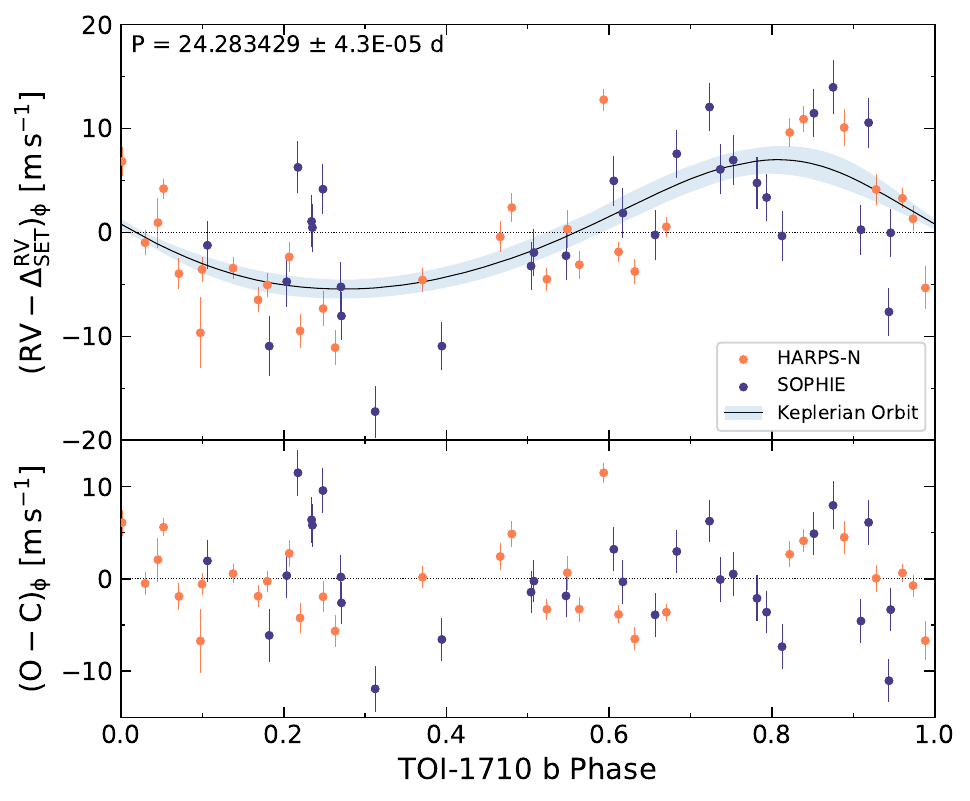}
    \caption{Phase-folded RV evolution of TOI-1710 determined from the  SOPHIE (in blue) and the HARPS-N (in orange) observations. The top panel shows the RV variation relative to a fitted instrumental offset (see text for details). The solid line and its light blue overlay correspond to the orbital solution presented in this study (Sect.~\ref{sec:models_parametrization_without_GPs}) with its relative 1$\upsigma$ uncertainty, respectively, and used to compute the residuals in the bottom panel.}\label{fig:RV_phase}
\end{figure}

\begin{table}
\newcolumntype{C}{ @{}>{{}}c<{{}}@{} }
\centering
\caption[]{Best-fit parameters for TOI-1710\:b (see Sect.~\ref{sec:results}).}\label{tab:best_fit}
\begin{tabular}{llrCl}
    \toprule
    Parameter & &  \multicolumn{3}{c}{Value (±1$\upsigma$)} \\ 
    \hline
    \multicolumn{2}{l}{\footnotesize{\underline{Stellar Parameters}$\mathrm{^{*}}$}} & & \rule{0pt}{2.6ex} \\ 
    $\rho_{\rm \star}$ & $\mathrm{[g\,cm^{-3}]}$   & $1.53$ & $\:\pm\:$ & $0.16$ \\
    $(u_{\rm \star},\:v_{\rm \star})$ &     & $(0.44,\:0.05)$ & $\:\pm\:$ & $(0.15,\:0.23)$ \\
    
    \multicolumn{2}{l}{\footnotesize{\underline{Instrument Offsets}}} \rule{0pt}{2.6ex} \\ 
    $\braket{RV}_{\textsf{HARPS-N}}$ & $\mathrm{[km\,s^{-1}]}$    & $-38.8171$ & $\:\pm\:$ & $0.8\:10^{-3}$ \\
    $\braket{RV}_{\textsf{SOPHIE}}$ & $\mathrm{[km\,s^{-1}]}$ & $-38.8549$ & $\:\pm\:$ & $1.1\:10^{-3}$ \\
    \multicolumn{2}{l}{\footnotesize{\underline{Planetary Parameters}}} & & \rule{0pt}{2.6ex} \\ 
    $T_{0}$ & $\mathrm{[BJD]}$    & $2459031.23025$ & $\:\pm\:$ & $4.2\:10^{-4}$ \\
    $P$ & $\mathrm{[d]}$   & $24.283429$ & $\:\pm\:$ & $4.3\:10^{-5}$ \\
    $K$ & $\mathrm{[m\,s^{-1}]}$   & $6.4$ & $\:\pm\:$ & $1.0$ \\
    $\sqrt{e}\,\sin\omega$ &  & $0.10$ & $\:\pm\:$ & $0.23$ \\
    $\sqrt{e}\,\cos\omega$ &  & $0.32$ & $\:\pm\:$ & $0.14$ \\
    $R_{\rm p}/R_{\rm \star}$ &     & $5.06\,\%$ & $\:\pm\:$ & $0.06\,\%$ \\
    $a/R_{\rm \star}$ &     & $34.2$ & $\:\pm\:$ & $3.9$ \\
    $b$ &    & \multicolumn{3}{c}{$<0.36$} \\
    
    \multicolumn{2}{l}{\footnotesize{\underline{Derived Parameters}$\mathrm{^{*}}$}} \rule{0pt}{2.6ex} \\ 
    $T_{14}$ & $\mathrm{[h]}$   & $5.24$ & $\:\pm\:$ & $0.03$ \\
    $a$ & $\mathrm{[AU]}$   & $0.16$ & $\:\pm\:$ & $0.04$ \\
    $R_{\rm p}$ & $[{\rm R}_{\rm \earth}]$   & $5.34$ & $\:\pm\:$ & $0.11$ \\
    $M_{\rm p}$ & $[{\rm M}_{\rm \earth}]$   & $28.3$ & $\:\pm\:$ & $4.7$ \\
    $\rho_{\rm p}$ & $[\mathrm{g\,cm^{-3}}]$   & $0.94$ & $\:\pm\:$ & $0.22$ \\
    $e$ &  & $0.16$ & $\:\pm\:$ & $0.08$ \\
    $\omega$ & $\mathrm{[^{\circ}]}$   & $17$ & $\:\pm\:$ & $44$ \\
    $i$ & $\mathrm{[^{\circ}]}$   & $89.6$ & $\:\pm\:$ & $0.3$ \\
    $\mathrm{RMS}^{\rm res.}_{\textsf{TESS}}$ &  & \multicolumn{3}{c}{~~~~~$6.5\:10^{-4}$} \rule{0pt}{2.6ex} \\
    $\mathrm{RMS}^{\rm res.}_{\textsf{HARPS-N}}$ & $\mathrm{[m\,s^{-1}]}$   & \multicolumn{3}{c}{~~~~~$4.1$}\\
    $\mathrm{RMS}^{\rm res.}_{\textsf{SOPHIE}}$ & $\mathrm{[m\,s^{-1}]}$   & \multicolumn{3}{c}{~~~~~$5.7$} \\
    
    \multicolumn{2}{l}{\footnotesize{\underline{Jitter amplitude (white noise)}}} \rule{0pt}{2.6ex} \\ 
    $\sigma_{\textsf{TESS}}$ &  & $2.24\:10^{-4}$ & $\:\pm\:$ & $5\:10^{-6}$ \\
    $\sigma_{\textsf{HARPS-N}}$ & $\mathrm{[m\,s^{-1}]}$   & $4.1$ & $\:\pm\:$ & $0.7$ \\
    $\sigma_{\textsf{SOPHIE}}$ & $\mathrm{[m\,s^{-1}]}$   & $5.5$ & $\:\pm\:$ & $0.9$\rule[-1.2ex]{0pt}{0pt} \\

    \bottomrule
\end{tabular}
\color{Black}
\tablefoot{\normalsize $\mathrm{^{(*)}}$ Assuming stellar parameters obtained from the stellar analysis presented in Sect.~\ref{sec:star_prop} and Table~\ref{tab:star}: $M_{\rm \star}=0.984^{+0.050}_{-0.059}~{\rm M}_{\rm \sun}$, $R_{\rm \star}=0.968^{+0.016}_{-0.014}~{\rm R}_{\rm \sun}$, $\rho_{\rm \star}=1.53^{+0.15}_{-0.16}~{\rm g\:cm^{-3}}$, and $(u_{\rm \star},\:v_{\rm \star})=(0.6313,\:0.0827)\pm(0.0027,\:0.0025)$. We adopt these values because they are more accurate.}
\end{table}

The posterior distributions obtained from the samples using the MCMC algorithm allowed us to derive the maximum a posteriori estimate for each parameter, its final median value, and the standard deviation. The medians of the posterior distributions were evaluated as the final parameters and their $\pm$34.1\% quantiles were adopted as the associated 1$\upsigma$ uncertainties. The complete results of the fitted and derived parameters are listed in Table~\ref{tab:best_fit}. The phase-folded transit best-fit model is shown in Fig.~\ref{fig:LC_phase} along with the corresponding photometric residuals. Our RV best-fit model is shown in Fig.~\ref{fig:RV_phase} along with the corresponding RV residuals. 

The Keplerian oscillation is measured with an orbital period $P$ of $24.283429\pm0.000043$~days and a RV semi-amplitude $K$ at $6.4\pm1.0~\mathrm{m\,s^{-1}}$. The 16\% error on this latter value constitutes the main contribution to the uncertainty calculated for the mass of the planet $M_{\rm p}$, consequently also equal to 16\%. The RV residuals show root mean square (RMS) values of $\mathrm{RMS}^{\rm res.}_{\textsf{HARPS-N}}=4.1~\mathrm{m\,s^{-1}}$ and $\mathrm{RMS}^{\rm res.}_{\textsf{SOPHIE}}=5.7~\mathrm{m\,s^{-1}}$. In other words, the observed RV residual dispersion is larger than the respective RV error median values of $\widetilde{\sigma_{\textsf{RV}}}=1.4$ and $2.4~\mathrm{m\,s^{-1}}$ for each spectrograph. This suggests that the data are indeed sensitive to additional signals, for example ~to stellar variability with a RV jitter amplitude of $\sim$5\:m/s. 
The LS periodogram of the RV residuals after this fit is plotted in the lower panel of Fig.~\ref{fig:GLS_periodogram_RV_res}. It no longer shows the main peak near 24~days corresponding to the planet, but it still shows a less significant peak near 30~days, which could be due to the stellar jitter. 
Tests of alternative models are presented in Sect.~\ref{sec:tests_and_validation}, showing the robustness of our results.

By combining $a/R_{\rm \star}=34.2\pm3.9$, $R_{\rm p}/R_{\rm \star}=5.06\%\pm0.06\%$, $b=0.22\pm0.14$, and $e=0.16\pm0.08$ from our best-fit model with the stellar parameters obtained in Sect.~\ref{sec:star_prop}, we derive that TOI-1710\:b has a radius $R_{\rm p}=5.34\pm0.11~{\rm R}_{\rm \earth}$, a mass $M_{\rm p}=28.3\pm4.7~{\rm M}_{\rm \earth}$, and thus a density $\rho_{\rm p}=0.94\pm0.22~\mathrm{g\,cm^{-3}}$. 
The eccentricity is consistent with zero at the current $2\text{}\upsigma$ precision (i.e., $e\leqslant0.26$), translating in a quasi-circular orbit for this planet. Hence, the orientation of the orbit with respect to the ascending node is difficult to constrain correctly, which explains the large error on the argument of periastron $\omega=17^{\circ}\pm44^{\circ}$. The posterior distribution obtained for $b$ only allows setting an upper limit $b<0.36$ on this parameter.

Furthermore, the adjustment of this joint model allows us to fit anew the stellar radius of TOI-1710 and the value we obtained $R_{\rm \star}=0.968\pm0.016~{\rm R}_{\rm \sun}$ is in good agreement with that given by the spectroscopic analysis. Fitting the transit profile also leads to an estimate of the stellar density with a value of $\rho_{\rm \star}=1.53\pm0.16~\mathrm{g\,cm^{-3}}$, which corresponds to the Sun's average density. Finally, the quadratic law limb-darkening coefficient fitted from the transit model $(u_{\rm \star},\:v_{\rm \star})=(0.44,\:0.05)\pm(0.14,\:0.23)$ falls within the $1\text{}\upsigma$ uncertainties in the values calculated using the \texttt{PyLDTk}\footnote{\texttt{\href{https://github.com/hpparvi/ldtk}{https://github.com/hpparvi/ldtk}}} Python toolkit (Sect.~\ref{sec:limb_darkening_profile}). Because they are more precise, we ultimately adopted the fundamental stellar parameters obtained from the analysis presented in Sect.~\ref{sec:star_prop}. 


\subsection{Alternative modeling and tests}
\label{sec:tests_and_validation}

The analysis presented in Sect.~\ref{sec:models_parametrization_without_GPs} might be affected by systematic effects due to the assumed hypotheses or the adopted methods.
We test here the robustness of our results when using alternative modeling techniques.

\subsubsection{Stellar activity}

The high dispersion levels on the RV residuals resulting from the analysis above might be due to stellar activity.
In addition, the orbital period of the planet is near the rotation period of its star, this suggests our derived planetary parameters might be affected.
Likewise, the photometric variations outside of the transits, although measured from the PDC-SAP flux mitigating long-term trends, possibly indicate the presence of stellar-variability perturbations.
To capture its quasi-periodic behavior, we chose to embed the same model presented above within a GP regression model \citep[\texttt{celerite2}\footnote{\texttt{\href{https://celerite2.readthedocs.io/en/latest/}{https://celerite2.readthedocs.io}}} v0.2.0,][]{2018RNAAS...2...31F} to probabilistically account for stellar magnetic activity perturbations and test their effects on the derived parameters.

We implemented three different GP kernels. First, we tested a stochastically driven harmonic oscillator kernel, statistically matching the quasi-periodic nature of the stellar correlated noise, as already proven successful (e.g., \citealt{2020A&A...634A..75B}). We also tried the simpler squared-exponential kernel and the flexible quasi-periodic kernel introduced by \citet{2017AJ....154..220F}, considered as drop-in replacements, for a significant gain in computational efficiency, of a quasi-periodic covariance function that has been proven to make valid probabilistic measurements even with sparsely sampled data \citep{2018MNRAS.474.2094A}. 

In each case the results (values and uncertainties) are in good agreement with those reported in Table~\ref{tab:best_fit}, obtained with the non-GP model, and that we finally adopted. Since testing alternative modelings explores possible effects of jitter on the retrieved planetary parameters of TOI-1710\:b, this reinforces their robustness.

\subsubsection{Alternative differential evolution MCMC modeling}

We also performed an independent joint analysis of HARPS-N and SOPHIE radial velocities and \textit{TESS} photometry, after normalizing the transits through local linear fitting, with a DE-MCMC method \citep{2013PASP..125...83E,2019arXiv190709480E}, following the same implementation as in \citet{2014A&A...572A...2B,2015A&A...575A..85B}. The obtained results are fully consistent (within 1$\upsigma$) with those reported in Table~\ref{tab:best_fit}.

\subsubsection{Model featuring a hypothetical second planet}

Finally, a simple two-planet model was tested on \texttt{juliet}\footnote{\texttt{\href{https://juliet.readthedocs.io/en/latest/}{https://juliet.readthedocs.io}}} \citep{2019MNRAS.490.2262E}, freezing both eccentricities to zero.
This allowed us to investigate the impact of a hypothetical second planet on our results.
It shows a slightly higher Bayesian log-evidence with respect to any other model ($\Delta\ln Z=1.5$, compared to the eccentric one-planet case and quasi-periodic kernel) which points to the possible presence of a lighter object with an orbital period of $30\pm0.4~\mathrm{days}$, causing a $3.8\pm0.9$ ms$^{-1}$ RV semi-amplitude $K$.
It is worth mentioning that adding a GP to this model did not increase its likelihood. Moreover, in this scenario the evaluated parameters of the first planet do not significantly differ from the one-planet model results, but the jitter levels are higher ($\sigma_{\textsf{HARPS-N}}=3.5\pm0.6~\mathrm{m\:s^{-1}}$, $\sigma_{\textsf{SOPHIE}}=5.6\pm0.8~\mathrm{m\:s^{-1}}$), as expected.
However, no transit of this hypothetical second planet has been observed in the \textit{TESS} LC, so the confirmation of a non-transiting planet would certainly require more evidence. We conclude that the results presented in Table~\ref{tab:best_fit} are robust with respect to the possibility of the presence of a second planet around TOI-1710.


\section{Discussion \& conclusions}
\label{sec:discussion_conclusions}

We report the discovery and the characterization of a warm super-Neptune transiting the bright ($V=9.6$) G-type star TOI-1710. The joint collection of HARPS-N and SOPHIE data has made possible a rapid spectroscopic follow-up, leading to evidence of its planetary nature and the characterization of the planet candidate detected by \textit{TESS}.

\subsection{The super-Neptunian gap}

\begin{figure}[ht]
    \centering
    \includegraphics[trim=0.0cm 0.0cm 0.0cm 0.0cm, clip=true, width=0.49\textwidth]{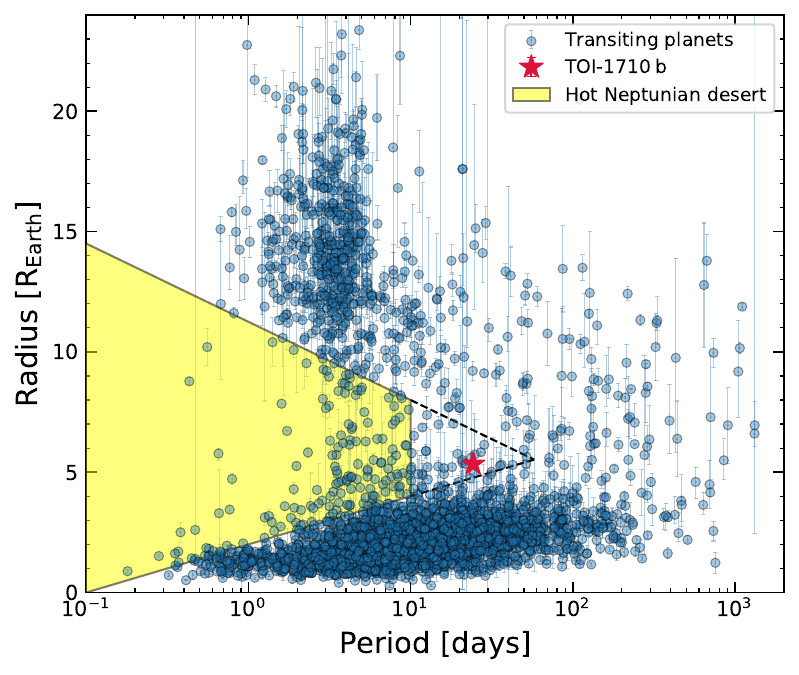}
    \caption{Planetary radius {vs.} orbital period diagram of the known transiting planets from the \texttt{exoplanet.eu} archive. The approximate expanse of the  hot-Neptune desert is highlighted in yellow and the adjacent dashed lines show the stretching boundaries including the warm Neptunes.}
    \label{fig:P_R_diagram}
\end{figure}

The term super-Neptune is used to refer to planets that are more massive than Neptune and less massive than Saturn, with estimated masses in the range of 20--80$~{\rm M}_{\rm \earth}$, and generally described as around 5 to 7 times larger than the Earth. 
\citet{2014A&A...572A...2B} provide an example of this nomenclature, with the first fully characterized super-Neptune Kepler-101\:b. The planets falling within this mass range are also referred to as sub-Saturn, for example TOI-257\:b \citep{2021MNRAS.502.3704A} and HD\:332231\:b \citep{2021arXiv211010282S}.

Exoplanet searches have so far revealed two dominant size and mass categories, on the one hand small rocky planets, namely Earth- to super-Earth-sized planets, and on the other hand, gaseous giants, Jupiter-sized planets and beyond \citep{2011arXiv1109.2497M,2019AJ....158..109H}. 
The distribution of the known exoplanets thus shows two peaks around the corresponding masses, digging an apparent valley around the sizes  between those of Neptune and Saturn.
There have been relatively few discoveries of planets of this size. The observed mass gap between Neptune-like and Jupiter-like planets, the hot-Neptune desert, is interpreted to originate from runaway accretion occurring for protoplanets of more than $20~{\rm M}_{\rm \earth}$. Above this mass threshold enough disk material is expected to aggregate onto the planet, letting it reach the size of Jupiter and beyond \citep{2018ApJ...869L..34S}, depopulating the 20--80$~{\rm M}_{\rm \earth}$ mass range, and thus forming a transition category of planets between the super-Earth and gaseous giants.

The expanse covered by the desert is shown on the radius {vs.} period distribution plot in Fig.~\ref{fig:P_R_diagram}.
According to the characterization presented in this study, the planet TOI-1710\:b lies on the edge of the desert, with a mass $M_{\rm p}=28.3\pm4.7~{\rm M}_{\rm \earth}$ and a radius $R_{\rm p}=5.34\pm0.11~{\rm R}_{\rm \earth}$; these values indeed make TOI-1710\:b fall into this scarce category of super-Neptune.

\subsection{Mass-radius diagram and internal structure}
\label{sec:discussion_internal_structure}

\begin{figure}[htbp]
    \centering
    \includegraphics[trim=0.0cm 0.0cm 0.0cm 0.0cm, clip=true, width=0.49\textwidth]{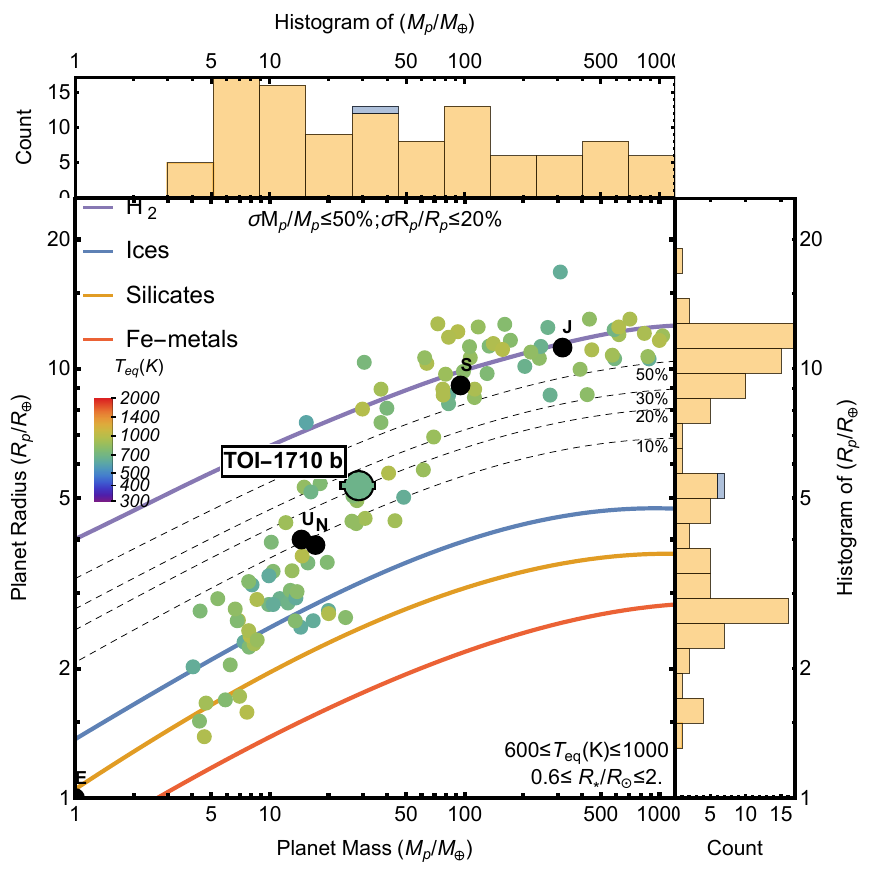}
    \caption{Planetary masses and radii (\textit{main panel}), and histogram of the masses (\textit{top panel}) and radii (\textit{right panel}) of the known transiting exoplanets \citep[TepCat catalog,][]{2011MNRAS.417.2166S} with equilibrium temperature $T_{\rm eq}$ evaluated between 600 and 1000~K. The composition models of hydrogen-, ice-, silicate-, and Fe-metal-rich planets (solid colored lines) are computed from \citet{2007ApJ...669.1279S}, \citet{2013PASP..125..227Z}, and \citet{2014ApJS..215...21B} (see details in Sect.~\ref{sec:discussion_internal_structure}). The dashed lines show the composition models with a relatively cold hydrogen envelope of mass fractions of 10\%, 20\%, 30\%, and 50\%. The Solar System planets are shown as gray dots.}
    \label{fig:combined_M_R_diagram}
\end{figure}

In order to compare TOI-1710\:b to the other known exoplanets of its range, we selected the ones from the TepCat\footnote{\texttt{\href{https://www.astro.keele.ac.uk/jkt/tepcat/}{https://www.astro.keele.ac.uk/jkt/tepcat/}}} catalog \citep[][]{2011MNRAS.417.2166S} orbiting solar-sized host stars ($0.6\leqslant R_{\rm \star}/{\rm R}_{\rm \sun}\leqslant2$) with equilibrium temperatures $T_{\rm eq}$ comprised between 600~K and 1000~K and relative mass and radius precisions below 50\% and 20\%, respectively. 

As can be seen in the exoplanetary mass-radius (M-R) diagram in Fig.~\ref{fig:combined_M_R_diagram}, TOI-1710\:b is not an outlier; it lies along the main sequence of exoplanets, probably resulting from intrinsic conformity in the composition resulting from the equations of state and formation processes of exoplanets \citep{2014ApJS..215...21B}. A further argument that can be made from this observation is that the super-Neptune TOI-1710\:b lies slightly beyond Uranus and Neptune, both in mass and radius, right below the domain of giant gaseous planets. 

In Fig.~\ref{fig:combined_M_R_diagram} the red, yellow, and blue lines respectively correspond to the mass-radius curves for pure-Fe metals, pure silicates, and high-pressure ices core compositions.
The purple line and the underlying dashed lines represent a pure-$\mathrm{H_2}$ composition and the corresponding models with a relatively cold hydrogen envelope of mass fractions of 10\%, 20\%, 30\%, and 50\% \citep[computed from][]{2014ApJS..215...21B}. These contours are approximately identical regardless of the core composition, rocky or icy.
The theoretical details, codes, and motivations of these composition models are explained by \citet{2021ApJ...923..247Z}.

The planetary density of TOI-1710\:b $\rho_{\rm p}=0.94\pm0.22~\mathrm{g\,cm^{-3}}$ lies significantly below Neptune’s mean density. In terms of its own composition, it likely has a somewhat more gaseous envelope compared to Uranus and Neptune; according to its place in the M-R diagram in Fig.~\ref{fig:combined_M_R_diagram}, its gaseous envelope is about $20\%$ to $30\%$ of its mass fraction, with an ice-rich core. In comparison, Uranus and Neptune each have about a $10\%$ envelope.
It appears that the radius of TOI-1710\:b is not sensitive to its core composition, whether rocky or icy, because of the presence of an extensive envelope, and the density difference between its core and envelope is significant.
There is however a degeneracy to be mentioned: the presence of core materials and heavy elements in the core has a similar effect on the planet radius as the same amount of material mixed throughout the envelope. The presence of a completely differentiated core, as distinct from just heavy element enrichment in the envelope, thereby remains unresolved by looking at the planetary mass and radius (and thus its mean density) alone. 
Therefore, the core materials may be partially mixed into the envelope to a certain extent. This has been suggested for Jupiter by the Juno mission \citep[see][and references therein]{Liu_et_al_2019}. 
For instance, some interior models of Uranus and Neptune also suggest that their interiors are partially mixed, and thus a gradual compositional gradient exists in the radial direction instead of a clear-cut core--envelope boundary \citep[see][and references therein]{2020SSRv..216...38H}. 

Finally, the details of the histograms presented in Fig.~\ref{fig:combined_M_R_diagram} show the relative dearth of super-Neptune planets immediately above TOI-1710\:b, between 6 and 7$~{\rm R}_{\rm \earth}$ in radii and between 40 and 70$~{\rm M}_{\rm \earth}$ in masses, for exoplanets with equilibrium temperatures below 1000~K. The fact that TOI-1710\:b lies right at the lower edge of this observed gap means that it is an interesting object to study the super-Neptunian transition between small planets and gaseous giants.

\subsection{Potentially non-zero eccentricity and formation scenario}

\begin{figure}[ht]
    \centering
    \includegraphics[trim=0.0cm 0.0cm 0.0cm 0.0cm, clip=true, width=0.49\textwidth]{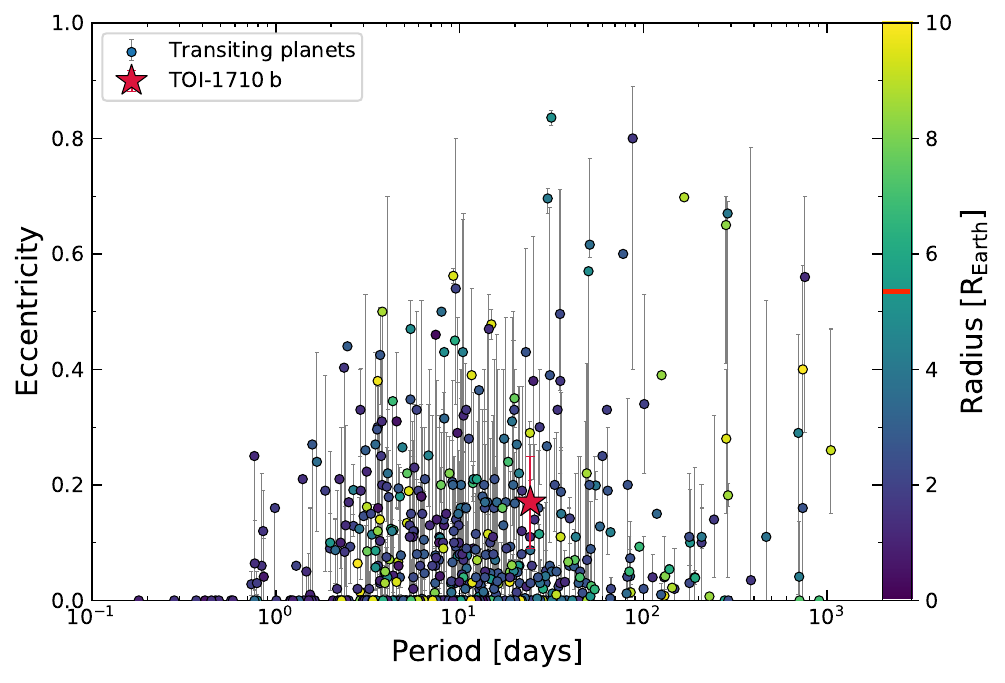}
    \caption{Orbital eccentricity (and uncertainties) {vs.} orbital period diagram of the known transiting planets from the \texttt{exoplanet.eu} archive with their planetary radii in color scale.}
    \label{fig:P_e_diagram}
\end{figure}

The photometric and spectroscopic joint fit reported in this study gives a near zero, but potentially non-zero, eccentricity of $e=0.16\pm0.08$. Resulting from the MCMC convergence, this parameter is distributed with 90\% of the posterior values between $0.04$ and $0.30$. As shown in the ($P$,\:$e$) diagram in Fig.~\ref{fig:P_e_diagram}, TOI-1710\:b might be part of the few eccentric planets known in that range of radii ($\leqslant10~{\rm R}_{\rm \earth}$) and orbital periods. So even though the best-fit value of $e$ is compatible with zero by 2$\upsigma$ and the conservative conjecture of a circular orbit remains valid, it appears legitimate to also investigate an eccentric scenario. Even so, it should be noted that low eccentricities can be overestimated by this kind of model, as shown by \citet{2013PASP..125...83E}, among others, and more fundamentally by \citet{1971AJ.....76..544L}.

A non-zero eccentricity would indicate the possibility of a non-in situ formation or migration process of this planet. For instance, it could result from gravitational perturbations, such as planet-planet scattering or a giant impact event during its formational stage. In particular, it could have formed from the merging of two or more smaller planetary bodies. 
A suspected example of such a scenario in our Solar System is Uranus; its eccentricity of 0.05 and the fact that it is spinning on its side with a retrograde rotation are usually interpreted as resulting from a giant impact event that occurred early in the lifetime of the planet \citep{2018ApJ...861...52K}.

\subsection{Prospect for atmospheric characterization via transmission spectroscopy}
\label{sec:transmission_spectroscopy}

\begin{figure}[ht]
    \centering
    \includegraphics[trim=0.0cm 0.0cm 0.0cm 0.0cm, clip=true, width=0.49\textwidth]{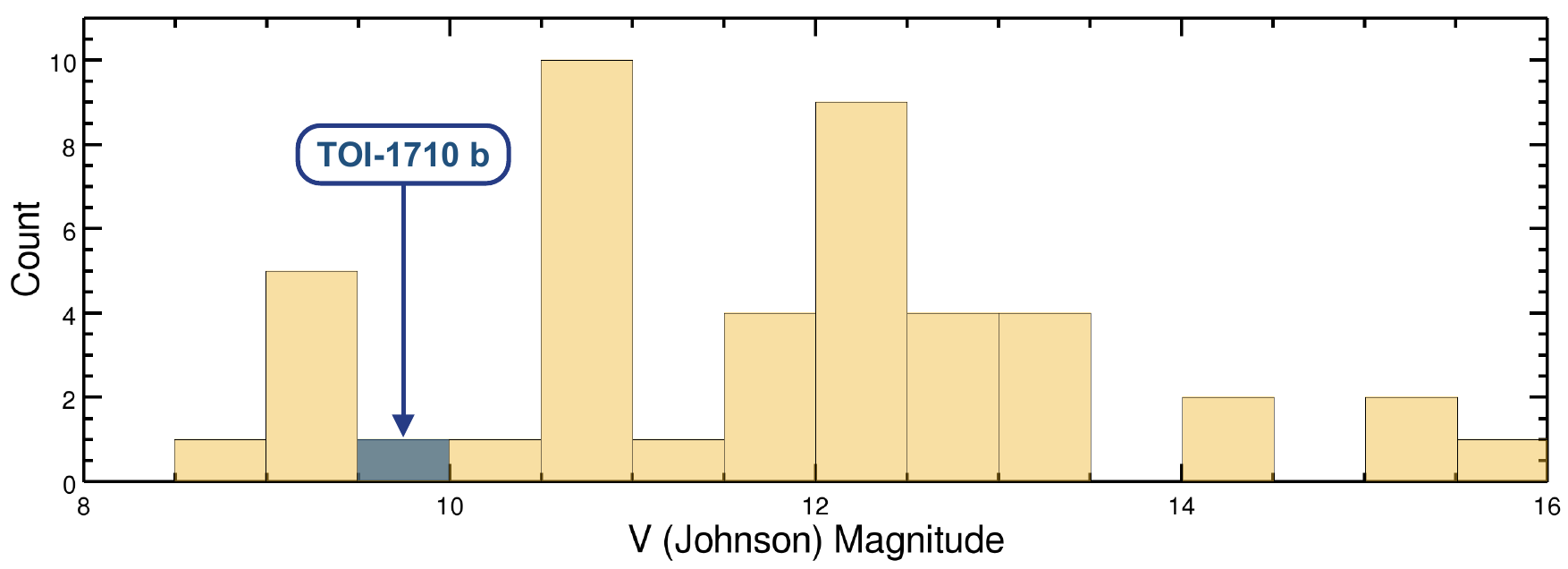}
    \caption{Histogram of the visible magnitudes of the host stars of the known transiting exoplanets with planetary radii and masses in the same range as TOI-1710\:b ($4\leqslant R_{\rm p}/{\rm R}_{\rm \earth}\leqslant7$ and $20\leqslant M_{\rm p}/{\rm M}_{\rm \earth}\leqslant40$) from the NASA exoplanet archive.}
    \label{fig:histogram_mV}
\end{figure}

In order to compare TOI-1710\:b with other planets, we selected the subgroup of planets with radii $4\leqslant R_{\rm p}/{\rm R}_{\rm \earth}\leqslant7$ and masses $20\leqslant M_{\rm p}/{\rm M}_{\rm \earth}\leqslant40$ that are transiting a $V<10$ star (see Fig.~\ref{fig:histogram_mV}). This group comprises the small planets among the transitional planets (sub-Jovian category, as defined by \citealt{2018PASP..130k4401K}) with about a 2$\upsigma$ mass range around that of TOI-1710\:b; TOI-1710\:b is one of those hosted by the brightest stars.

We discuss here the possibility of further characterization of the planet, in particular by examining the potential of \textit{JWST} in-transit observations to detect the presence of molecular features in transmission spectra. We calculate a transmission spectroscopy metric (TSM) \citep[cf.][Eq.~1]{2018PASP..130k4401K} of 97, with an equilibrium temperature $T_{\rm eq}=687\pm~50\mathrm{K}$, placing TOI-1710\:b on a second quartile rank for transmission spectroscopy follow-up efforts, and making it a good candidate for atmospheric characterization. Therefore, future follow-up observations will not only allow the search for additional planets in the TOI-1710 system, but might also help constrain low-mass planet formation and evolution models, key to understanding the hot-Neptune desert.

\begin{figure}[htbp]
    \centering
    \includegraphics[trim=0.0cm 0.0cm 0.0cm 0.0cm, clip=true, width=0.49\textwidth]{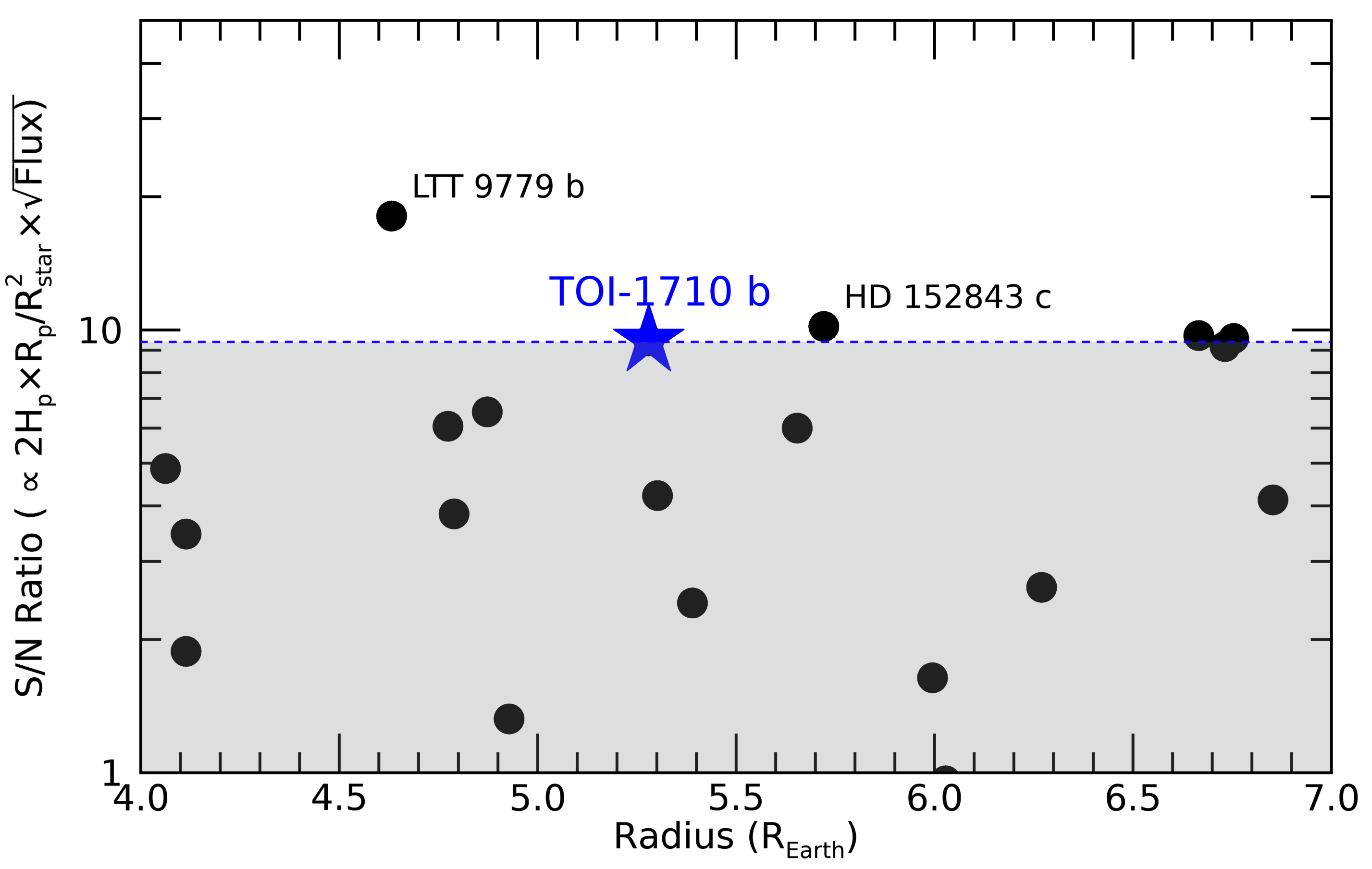}
    \caption{Predicted S/N for transmission spectroscopy observations in the V band of the known transiting exoplanets from the \texttt{exoplanet.eu} archive orbiting G-dwarf (or $5000\leqslant T_{\rm eff}\leqslant6000~\mathrm{K}$) stars, and with planetary radii and masses in the same range as TOI-1710\:b ($4\leqslant R_{\rm p}/{\rm R}_{\rm \earth}\leqslant7$ and $20\leqslant M_{\rm p}/{\rm M}_{\rm \earth}\leqslant40$) {vs.} planetary radii. The S/N shown is normalized to 100 for HD\:209458\:b.}
    \label{fig:SNR_transmission_vs_radius}
\end{figure}

We gauged the pertinence of TOI-1710\:b as a target for atmospheric detection in the optical band with respect to the planets in the same range. More specifically, we selected a class of the known transiting exoplanets orbiting G-dwarf stars (or with effective temperatures between $5000$ and $6000~\mathrm{K}$ when the spectral type of the host star is unknown) and among them, those that are in the same size and mass ranges as TOI-1710\:b ($4\leqslant R_{\rm p}/{\rm R}_{\rm \earth}\leqslant7$ and $20\leqslant M_{\rm p}/{\rm M}_{\rm \earth}\leqslant40$). The predicted S/N for observations in transmission spectroscopy in the V band can be calculated from the relation
\begin{equation}
    S/N \propto 2 H_{\rm p} \times \dfrac{R_{\rm p}}{R_{\rm star}} \times \sqrt{\mathrm{Flux}_{\rm V}},
\end{equation}
where $H_{\rm p}$ represents the atmospheric scale height of a planet and $\mathrm{Flux}_{\rm V}$ the flux of its host star in the V band. We normalized this ratio to 100 for the optimal target HD\:209458\:b. Plotting the S/N {versus} the radii of this selection of exoplanets (see Fig.~\ref{fig:SNR_transmission_vs_radius}) ranks TOI-1710\:b at the fourth position with a predicted S/N just above 9, which means that TOI-1710\:b is a good candidate for atmospheric characterization via transmission spectroscopy and figures among the top tier targets of its class in the optical band. This could result in a potential detection of sodium or potassium absorption lines in the atmosphere of this planet if the spectral features are not hidden by haze absorption \citep{2016Natur.529...59S}; in this regard, TOI-1710\:b provides a new target for characterizing the physical condition for the formation of clouds.

\begin{acknowledgements}
We thank the anonymous referees for their useful comments.
Funding for this study was provided by ESO and IAP. 
SOPHIE spectroscopic data were secured at Observatoire de Haute-Provence, France. 
HARPS-N spectroscopic data were secured at the Roque de los Muchachos Observatory, La Palma, Spain. 
This work makes use of observations from the LCOGT network. Part of the LCOGT telescope time was granted by NOIRLab through the Mid-Scale Innovations Program (MSIP). MSIP is funded by NSF.
This paper includes data collected by the \textit{TESS} mission, which are publicly available from the Mikulski Archive for Space Telescopes (MAST). Funding for the \textit{TESS} mission is provided by NASA’s Science Mission directorate. 
We acknowledge the use of public \textit{TESS} data from pipelines at the \textit{TESS} Science Office and at the \textit{TESS} Science Processing Operations Center. 
Resources supporting this work were provided by the NASA High-End Computing (HEC) Program through the NASA Advanced Supercomputing (NAS) Division at Ames Research Center for the production of the SPOC data products.
We acknowledge the support by FCT~-- Fundação para a Ciência e a Tecnologia through national funds and by FEDER through COMPETE2020~-- Programa Operacional Competitividade e Internacionalização by these grants: UID/FIS/04434/2019; UIDB/04434/2020; UIDP/04434/2020; PTDC/FIS-AST/32113/2017 \& POCI-01-0145-FEDER-032113; PTDC/FISAST/28953/2017 \& POCI-01-0145-FEDER-028953. This research has made use of data obtained from the portal \texttt{exoplanet.eu} of The Extrasolar Planets Encyclopaedia. This work has made use of data from the European Space Agency (ESA) mission {\it Gaia} (\url{https://www.cosmos.esa.int/gaia}), processed by the {\it Gaia} Data Processing and Analysis Consortium (DPAC, \url{https://www.cosmos.esa.int/web/gaia/dpac/consortium}). Funding for the DPAC has been provided by national institutions, in particular the institutions participating in the {\it Gaia} Multilateral Agreement. 
P.~Cortés-Zuleta thanks the LSSTC Data Science Fellowship Program, which is funded by LSSTC, NSF Cybertraining Grant \#1829740, the Brinson Foundation, and the Moore Foundation; her participation in the program has benefited this work. G.~Lacedelli acknowledges support by CARIPARO Foundation, according to the agreement CARIPARO-Università degli Studi di Padova (Pratica n.~2018/0098). P.-C.~König acknowledges the support from the IMPRS on Astrophysics, Garching, Germany. M.Damasso acknowledges financial support from the FP7-SPACE Project ETAEARTH (GA no. 313014).
The author L. Zeng would like to thank the support by the U.S. Department of Energy under awards DE-NA0003904 and DE-FOA0002633 (to S.B.J., principal investigator, and collaborator L. Zeng) with Harvard University and by the Sandia Z Fundamental Science Program. This research represents the authors’ views and not those of the Department of Energy.
The authors G. Hébrard and I. Boisse acknowledge the support from the French Programme National de Planétologie (PNP) of CNRS (INSU).
\end{acknowledgements}


\begin{balance}
\bibliographystyle{aa} 
\bibliography{AA_2021_43002.bib}
\end{balance}



\begin{appendix}


\section{\textit{TESS} apertures}
\label{appendix:TESS_apertures}

\begin{figure}[hbtp]
\centering
\includegraphics[trim=0cm 0cm 0cm 0cm, clip=true, scale=0.77, angle =0]{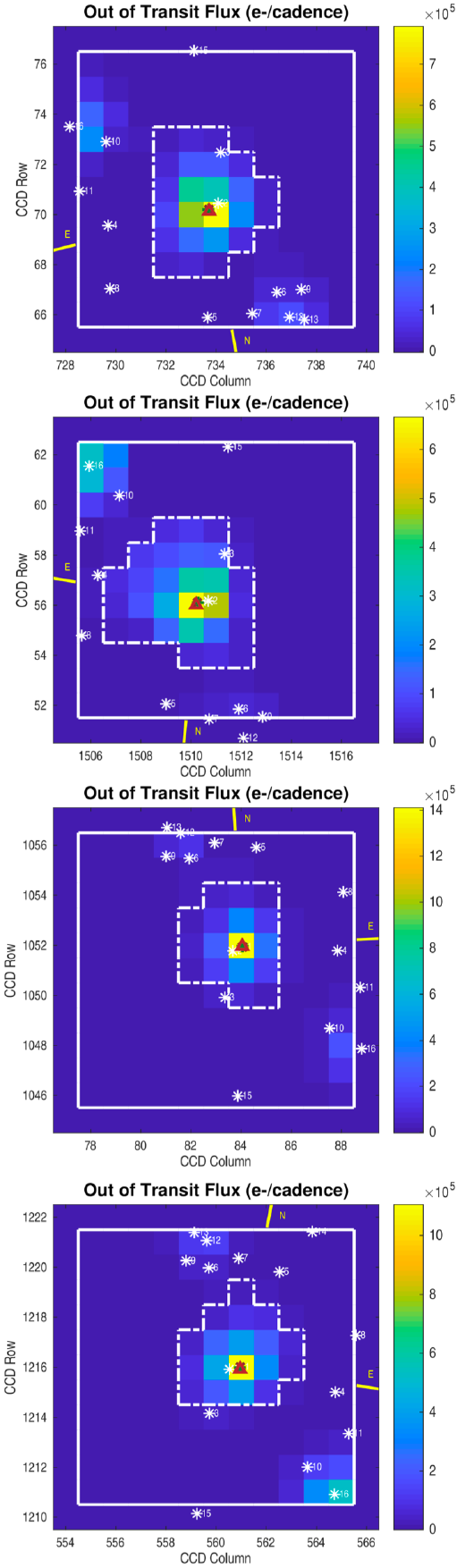}
\caption{\textit{TESS} images of TOI-1710 out of transit, as provided by the SPOC pipeline Data Validation Reports of 2 August, 2020, and 21 August 2021, in Sectors 19, 20, 26, and 40 (from top to bottom). \textit{TESS} pixels cover a sky aperture of 21~arcsec.}\label{fig:TESS_apertures}
\end{figure}

The \textit{TESS} full images in the vicinity of the targets TOI-1710 are shown in Fig.~\ref{fig:TESS_apertures}. The pixels enclosed within the dashed white contours are those used in the aperture for the PDC-SAP flux. The targeted star has faint neighbours, at distances large enough so that the contamination of the main target apertures is not affected; its closest neighbour is located at a distance of 8.69~arcsec with a \textit{TESS} magnitude of 17.72.

\section{Prior and posterior distributions}
\label{appendix:prior}

\begin{figure*}[ht]
\centering
\includegraphics[trim=0cm 0cm 0cm 0cm, clip=true, width=0.99\textwidth]{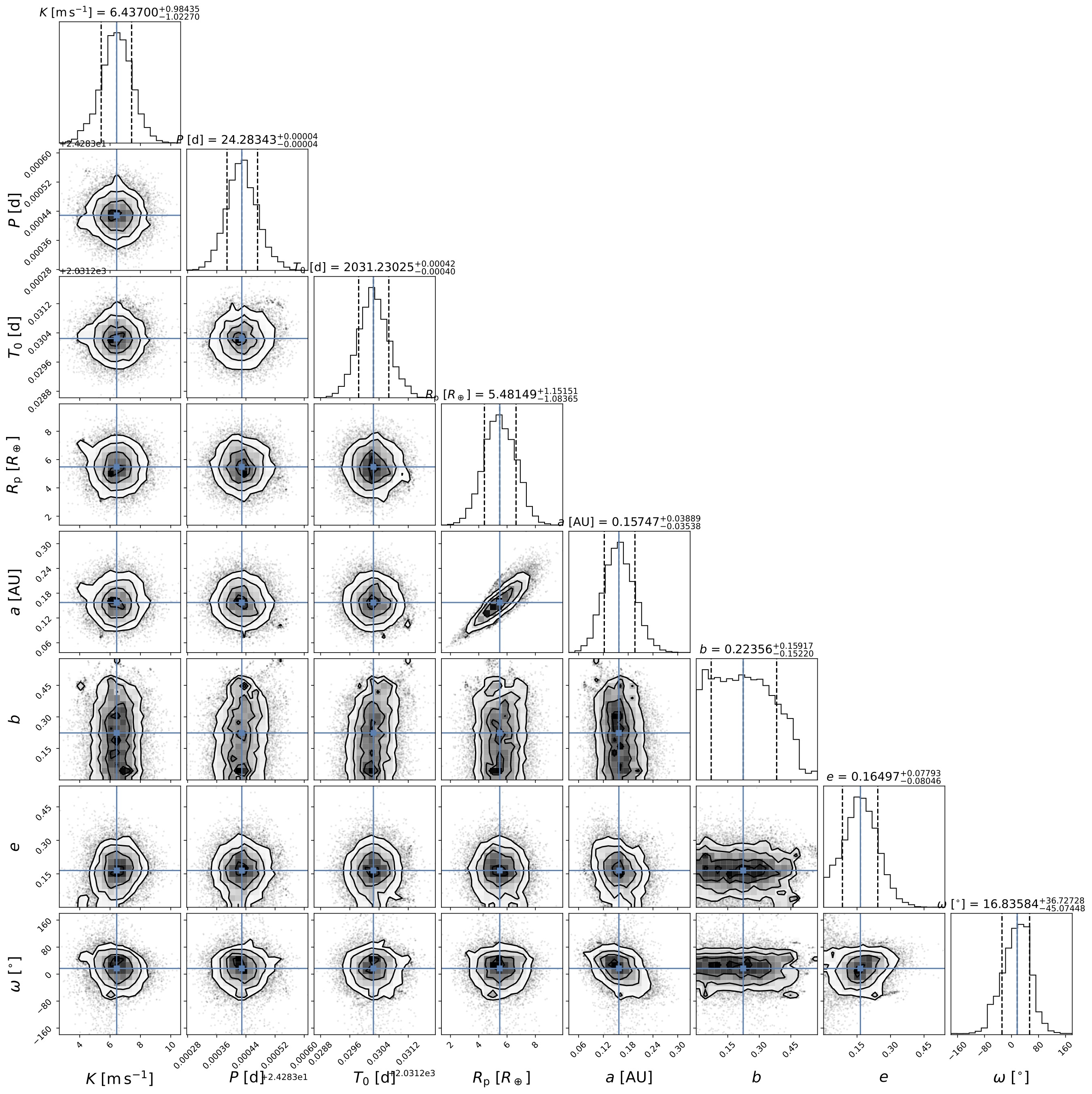}
\caption{Corner plot for the posterior distribution of the parameters of the transit profile and the RV eccentric orbit of TOI-1710 resulting from the MCMC fit (see Sect.~\ref{sec:models_parametrization_without_GPs}).}\label{fig:corner}
\end{figure*}

We list in Table~\ref{tab:prior} the prior distributions of the parameters of the model presented in Sect.~\ref{sec:transit_keplerian_characterization} and in Fig.~\ref{fig:corner} the corner plot of the posterior distributions resulting from our best MCMC fit.


\begin{table}[hbtp]
\centering
\caption{Prior parameter distributions of the global fit with the \texttt{PyMC3} and \texttt{celerite2} packages (see Sections~\ref{sec:models_parametrization_without_GPs}); $\mathcal{U}(a,\:b)$ indicates a uniform distribution between $a$ and $b$; $\mathcal{N}(a,\:b)$ a normal distribution with mean $a$ and standard deviation~$b$.}\label{tab:prior}
\begin{tabular}{llc}
    \toprule
    \multicolumn{2}{l}{Parameter} &  Prior distribution \\ 
    \hline
    \multicolumn{2}{l}{\footnotesize{\underline{Stellar Parameters}}} & \rule{0pt}{2.6ex} \\ 
    $R_{\rm \star}$ & $[{\rm R}_{\rm \sun}]$ & $\mathcal{N}(0.968,\:0.016)$ \\
    $(u_{\rm \star},\:v_{\rm \star})$ & & \citet{2013MNRAS.435.2152K}$^{\text{\textasteriskcentered}}$ \\
    
    \multicolumn{2}{l}{\footnotesize{\underline{Instrument Offsets}}} \rule{0pt}{2.6ex} \\ 
    $\braket{RV}_{\textsf{HARPS-N}}$ & $\mathrm{[km\,s^{-1}]}$ & $\mathcal{N}(-38.8,\:1.0)$ \\
    $\braket{RV}_{\textsf{SOPHIE}}$ & $\mathrm{[km\,s^{-1}]}$ & $\mathcal{N}(-38.8,\:1.0)$ \\
    \multicolumn{2}{l}{\footnotesize{\underline{Keplerian Parameters}}} & \rule{0pt}{2.6ex} \\ 
    $T_{0}$ & $\mathrm{[BJD]}$    & $\mathcal{N}(2459031.2,\:1.0)$ \\
    $\ln P$ &    & $\mathcal{N}(3.2,\:0.1)$ \\
    $\ln K$ &    & $\mathcal{U}(0,\:2.7)$ \\
    $\sqrt{e}\,\sin\omega$ &  & $\mathcal{U}(-1,\:1)$ \\
    $\sqrt{e}\,\cos\omega$ &  & $\mathcal{U}(-1,\:1)$ \\
    
    \multicolumn{2}{l}{\footnotesize{\underline{Transit Parameters}}} & \rule{0pt}{2.6ex} \\ 
    $R_{\rm p}/R_{\rm \star}$ &     & $\mathcal{N}(0.0529,\:0.010)$ \\
    $a/R_{\rm \star}$ &     & $\mathcal{N}(37,\:10)$ \\
    $b$ &    & $\mathcal{U}(0,\:1.053)$ \\
    \bottomrule
\end{tabular}
\tablefoot{$\mathrm{^{(\text{\textasteriskcentered})}}$ Implementation of the \citet{2013MNRAS.435.2152K} reparameterization of the two-parameter limb-darkening model to allow for efficient and uninformative prior sampling.}
\end{table}

\section{SOPHIE and HARPS-N spectroscopy}
\label{appendix:RVs}

We report in Table~\ref{tab:SOPHIE_RV} and Table~\ref{tab:HARPSN_RV} the 30+31 RV data points with their corresponding uncertainties, FWHM, BIS, exposure times, S/N, and spectral indices. 


\begin{table*}[b]
\centering
\caption[]{SOPHIE spectroscopic data obtained between 14 September 2020 and 7 May 2021.}\label{tab:SOPHIE_RV}
{\scriptsize
\begin{tabular}{lccccccc}
    \toprule
    $\mathrm{BJD_{\textsf{UTC}}}$ & RV & ±$1\upsigma_{\textsf{RV}}$ & FWHM & BIS$^{(\dagger)}$ & Exp.$^{(*)}$ & $\mathrm{S/N^{(\ddagger)}}$ & $\log R^{\prime}_{\rm HK}$ \\ 
    $-2\:450\:000.0$ & \multicolumn{2}{c}{$\mathrm{~~~[m\,s^{-1}]}$} & $\mathrm{[m\,s^{-1}]}$ & $\mathrm{[m\,s^{-1}]}$ & $\mathrm{[sec]}$ & -- & $\mathrm{[dex]}$\rule[-1.2ex]{0pt}{0pt} \\ 
    \hline
    $9106.66165$ & $-38856.1$ & $2.3$ & $7820$ & $-26$ & $540.8$ & $51.0$ & $-4.73$\rule{0pt}{2.6ex} \\ 
    $9110.66029$ & $-38862.9$ & $2.3$ & $7790$ & $-13$ & $848.7$ & $50.9$ & $-4.83$ \\
    $9111.66654$ & $-38872.1$ & $2.5$ & $7790$ & $-25$ & $1000.0$ & $47.9$ & $-4.75$ \\
    $9113.65844$ & $-38865.8$ & $2.3$ & $7770$ & $-21$ & $929.4$ & $50.4$ & $-4.80$ \\
    $9120.66839$ & $-38847.3$ & $2.3$ & $7790$ & $-36$ & $999.4$ & $50.6$ & $-4.77$ \\
    $9140.60268$ & $-38858.1$ & $2.3$ & $7780$ & $-16$ & $565.5$ & $50.7$ & $-4.86$ \\
    $9140.68790$ & $-38856.8$ & $2.3$ & $7790$ & $-27$ & $735.4$ & $50.7$ & $-4.72$ \\
    $9141.64650$ & $-38857.1$ & $2.3$ & $7790$ & $-28$ & $545.5$ & $50.6$ & $-4.84$ \\
    $9146.63540$ & $-38847.9$ & $2.4$ & $7800$ & $-22$ & $763.9$ & $50.5$ & $-4.64$ \\
    $9147.63244$ & $-38851.5$ & $2.3$ & $7800$ & $-12$ & $653.8$ & $50.6$ & $-4.85$ \\
    $9149.61554$ & $-38840.9$ & $2.6$ & $7810$ & $-20$ & $1000.0$ & $44.6$ & $-4.79$ \\
    $9150.67614$ & $-38844.3$ & $2.4$ & $7810$ & $-5$ & $1290.3$ & $50.3$ & $-4.69$ \\
    $9167.62705$ & $-38853.0$ & $2.4$ & $7800$ & $-21$ & $1702.3$ & $49.3$ & $-4.73$ \\
    $9168.58774$ & $-38855.1$ & $2.4$ & $7800$ & $-28$ & $669.0$ & $50.4$ & $-4.80$ \\
    $9170.53283$ & $-38848.8$ & $2.4$ & $7800$ & $-12$ & $817.4$ & $50.2$ & $-4.74$ \\
    $9175.56654$ & $-38862.5$ & $2.3$ & $7790$ & $-11$ & $614.8$ & $50.4$ & $-4.83$ \\
    $9182.64521$ & $-38854.4$ & $2.3$ & $7800$ & $-23$ & $480.2$ & $50.6$ & $-4.80$ \\
    $9183.49595$ & $-38860.1$ & $2.4$ & $7800$ & $-2$ & $611.7$ & $50.2$ & $-4.81$ \\
    $9194.49575$ & $-38842.8$ & $2.3$ & $7790$ & $-39$ & $579.5$ & $50.6$ & $-4.70$ \\
    $9197.61215$ & $-38843.4$ & $2.3$ & $7830$ & $-28$ & $539.7$ & $50.8$ & $-4.80$ \\
    $9206.49686$ & $-38848.6$ & $2.5$ & $7810$ & $-46$ & $1302.9$ & $48.1$ & $-4.77$ \\
    $9244.47901$ & $-38850.1$ & $2.5$ & $7800$ & $-25$ & $1000.0$ & $47.8$ & $-4.83$ \\
    $9247.58074$ & $-38854.6$ & $2.4$ & $7800$ & $-25$ & $911.8$ & $50.6$ & $-4.77$ \\
    $9248.46862$ & $-38854.9$ & $2.3$ & $7800$ & $-41$ & $537.0$ & $50.8$ & $-4.82$ \\
    $9278.48916$ & $-38865.8$ & $2.9$ & $7800$ & $-21$ & $1000.0$ & $41.3$ & $-4.84$ \\
    $9303.29701$ & $-38859.6$ & $2.4$ & $7800$ & $-11$ & $631.5$ & $50.6$ & $-4.88$ \\
    $9304.37681$ & $-38850.7$ & $2.4$ & $7830$ & $-8$ & $530.6$ & $50.8$ & $-4.85$ \\
    $9328.32128$ & $-38853.8$ & $2.5$ & $7800$ & $-27$ & $1000.0$ & $47.6$ & $-5.00$ \\
    $9337.34016$ & $-38849.9$ & $2.4$ & $7840$ & $-17$ & $1079.3$ & $50.7$ & $-4.88$ \\
    $9342.35977$ & $-38855.2$ & $2.4$ & $7830$ & $-8$ & $999.7$ & $51.2$ & $-4.71$ \rule[-0.8ex]{0pt}{0pt} \\

    \bottomrule
\end{tabular}}
\tablefoot{$^{(*)}$~Duration of each individual exposure. $^{(\dagger)}$~Bisector spans; error bars are twice those of RVs. $^{(\ddagger)}$~S/N per pixel at order 26, i.e., at $550\:\mathrm{nm}$.}
\end{table*}


\begin{table*}
\centering
\caption{HARPS-N spectroscopic data obtained between 3 October 2020 and 19 April 2021.}\label{tab:HARPSN_RV}
{\scriptsize
\begin{tabular}{lcccccccc}
    \toprule
    $\mathrm{BJD_{\textsf{UTC}}}$ & RV & ±$1\upsigma_{\textsf{RV}}$ & FWHM & BIS$^{(\dagger)}$ & Exp.$^{(*)}$ & $\mathrm{S/N^{\rm (\ddagger)}}$ & $I_{\rm Ca\:{\sc ii}}$ & $I_{\rm H\upalpha}$ \\ 
    $-2\:450\:000.0$ & \multicolumn{2}{c}{$\mathrm{~~~[m\,s^{-1}]}$} & $\mathrm{[m\,s^{-1}]}$ & $\mathrm{[m\,s^{-1}]}$ & $\mathrm{[sec]}$ & --& --& --\rule[-1.2ex]{0pt}{0pt} \\ 
    \hline
$9125.66393$ & $-38807.06$ & $1.79$ & $7190.2$ & $-13.3$ & $900.0$ & $55.0$ & $0.104$ & $0.107$\rule{0pt}{2.6ex} \\
$9126.61745$ & $-38813.01$ & $1.46$ & $7180.1$ & $-11.6$ & $1200.0$ & $66.1$ & $0.099$ & $0.109$ \\
$9127.71852$ & $-38815.83$ & $1.20$ & $7180.8$ & $-11.2$ & $900.0$ & $78.2$ & $0.099$ & $0.109$ \\
$9130.73516$ & $-38826.80$ & $3.42$ & $7173.5$ & $-23.6$ & $900.0$ & $32.8$ & $0.102$ & $0.109$ \\
$9133.71089$ & $-38826.62$ & $1.62$ & $7172.4$ & $-16.3$ & $900.0$ & $60.0$ & $0.094$ & $0.105$ \\
$9134.75545$ & $-38828.23$ & $1.72$ & $7170.8$ & $-18.0$ & $900.0$ & $57.1$ & $0.093$ & $0.108$ \\
$9156.74122$ & $-38823.64$ & $1.21$ & $7182.0$ & $-20.1$ & $900.0$ & $77.8$ & $0.097$ & $0.107$ \\
$9172.60465$ & $-38807.53$ & $1.40$ & $7180.4$ & $-11.4$ & $900.0$ & $68.5$ & $0.096$ & $0.107$ \\
$9188.59605$ & $-38814.75$ & $1.38$ & $7185.6$ & $-15.0$ & $900.0$ & $68.7$ & $0.100$ & $0.113$ \\
$9189.64237$ & $-38821.64$ & $1.12$ & $7189.3$ & $-13.1$ & $900.0$ & $81.6$ & $0.103$ & $0.108$ \\
$9190.60987$ & $-38820.25$ & $1.37$ & $7177.6$ & $-16.2$ & $900.0$ & $67.1$ & $0.101$ & $0.109$ \\
$9191.78142$ & $-38819.01$ & $0.98$ & $7181.0$ & $-17.9$ & $900.0$ & $97.1$ & $0.100$ & $0.108$ \\
$9212.53358$ & $-38817.57$ & $1.49$ & $7187.8$ & $-16.2$ & $900.0$ & $64.4$ & $0.106$ & $0.108$ \\
$9215.62219$ & $-38804.38$ & $1.08$ & $7189.0$ & $-20.8$ & $900.0$ & $86.3$ & $0.105$ & $0.109$ \\
$9275.33736$ & $-38812.94$ & $0.95$ & $7197.0$ & $-9.8$ & $900.0$ & $95.3$ & $0.106$ & $0.109$ \\
$9276.48561$ & $-38820.71$ & $1.22$ & $7182.6$ & $-11.7$ & $900.0$ & $76.5$ & $0.105$ & $0.114$ \\
$9277.41006$ & $-38820.58$ & $1.04$ & $7181.5$ & $-12.1$ & $599.4$ & $88.5$ & $0.102$ & $0.112$ \\
$9278.43231$ & $-38822.20$ & $1.14$ & $7182.5$ & $-14.9$ & $604.7$ & $81.6$ & $0.102$ & $0.108$ \\
$9287.38540$ & $-38816.83$ & $1.86$ & $7165.9$ & $-18.4$ & $900.0$ & $53.2$ & $0.103$ & $0.107$ \\
$9289.40035$ & $-38820.90$ & $1.24$ & $7171.5$ & $-17.3$ & $900.0$ & $75.7$ & $0.098$ & $0.110$ \\
$9290.34069$ & $-38816.60$ & $0.94$ & $7174.2$ & $-21.1$ & $900.0$ & $98.1$ & $0.099$ & $0.111$ \\
$9294.43466$ & $-38806.25$ & $1.23$ & $7198.1$ & $-19.6$ & $900.0$ & $77.3$ & $0.104$ & $0.112$ \\
$9297.38773$ & $-38813.86$ & $0.96$ & $7192.8$ & $-12.9$ & $900.0$ & $96.8$ & $0.106$ & $0.114$ \\
$9298.38306$ & $-38810.32$ & $1.39$ & $7187.4$ & $-13.5$ & $900.0$ & $68.9$ & $0.108$ & $0.114$ \\
$9299.45267$ & $-38816.21$ & $2.39$ & $7189.5$ & $-17.9$ & $617.4$ & $44.2$ & $0.106$ & $0.111$ \\
$9303.37054$ & $-38819.51$ & $1.49$ & $7177.6$ & $-16.6$ & $900.0$ & $65.3$ & $0.103$ & $0.106$ \\
$9304.38122$ & $-38824.48$ & $1.69$ & $7181.4$ & $-25.3$ & $900.0$ & $58.1$ & $0.097$ & $0.105$ \\
$9307.34891$ & $-38821.72$ & $1.18$ & $7171.6$ & $-18.8$ & $900.0$ & $80.9$ & $0.098$ & $0.105$ \\
$9322.35329$ & $-38822.48$ & $2.07$ & $7168.9$ & $-7.1$ & $900.0$ & $48.7$ & $0.103$ & $0.108$ \\
$9323.35210$ & $-38818.13$ & $1.22$ & $7175.8$ & $-20.5$ & $900.0$ & $76.0$ & $0.101$ & $0.105$ \\
$9324.36237$ & $-38821.12$ & $1.48$ & $7164.3$ & $-15.1$ & $900.0$ & $65.2$ & $0.102$ & $0.105$\rule[-0.8ex]{0pt}{0pt} \\

    \bottomrule
\end{tabular}}
\tablefoot{$^{(*)}$~Duration of each individual exposure. $^{(\dagger)}$~Bisector spans; error bars are twice those of RVs. $^{\rm (\ddagger)}$~S/N per pixel at order 50, i.e., at 564~nm.}
\end{table*}

\end{appendix}

\end{document}